\documentclass[amsmath,twocolumn,pra,notitlepage,footinbib]{revtex4-2}
\usepackage{amsthm,amsfonts,amssymb,graphicx,verbatim,xcolor,xfrac,bm}
\usepackage{hyperref,siunitx}
\usepackage[utf8]{inputenc}

\usepackage{tikz} 
\usetikzlibrary{arrows,decorations.pathmorphing,backgrounds,positioning,fit,matrix,calc}
\tikzset{snake it/.style={decorate, decoration=snake,segment length=0.1cm}}

\renewcommand{\vec}[1]{\bm{#1}}

\newcommand{\ket}[1]{|#1\rangle}

\newcommand{\braOket}[3]{\langle #1|#2|#3\rangle}

\begin{document}
\title{Anomalous Hall metal and fractional Chern insulator in \\ twisted transition metal dichalcogenides}
\author{Valentin Cr\'epel}
\author{Liang Fu}
\affiliation{Department of Physics, Massachusetts Institute of Technology, Cambridge, Massachusetts 02139, USA}

\begin{abstract}
We predict robust Ising ferromagnetism driven by Coulomb interaction in the metallic phase of twisted transition metal dichalcogenides homobilayers for a range of small twist angles. Due to the presence of spin-valley locking and Chern band, the completely spin polarized state---a half metal---has a spin gap and exhibits anomalous Hall effect.  We also find that near a magic angle where the Chern band is predicted to be exceptionally flat, the anomalous Hall metal at $1/3$ filling may become unstable  at low temperature to a $\sqrt{3}\times\sqrt{3}$ charge density wave, or a fractional Chern insulator.  
\end{abstract}

\maketitle

\paragraph*{Introduction ---} Moir\'e materials based on two-dimensional (2D) van der Waals heterostructures have swiftly become an exciting platform for the realization of strongly correlated and topological electronic phases. Among these artificial quantum materials, a particular class -- semiconductor transition metal dichalcogenide (TMD) heterobilayers -- has recently attracted great interest and intensive study due to a plethora of novel electronic phases discovered therein, including Mott-Hubbard and charge-transfer insulators~\cite{wu2018hubbard,zhang2020moire,tang2020simulation,regan2020mott}, generalized Wigner crystals~\cite{regan2020mott,xu2020correlated,huang2021correlated,li2021imaging}, and the quantum anomalous Hall (QAH) state~\cite{li2021quantum}.

Another 2D semiconductor based platform to search for correlated topological phases is  
twisted TMD homobilayers ($t$TMD). It was predicted early on~\cite{wu2019topological} that due to spin-valley locking and spatially modulated interlayer tunneling, $t$TMD host topological moir\'e bands of the Kane-Mele type~\cite{kane2005quantum} over a continuum of twist angles $\theta$. They would lead to a quantum spin Hall state at the filling of $n=2$ holes per moir\'e unit cell, which corresponds to the complete filling of the topmost moir\'e valence bands including both $K$ and $-K$ valleys. Even more interesting is the possibility that at half band filling $n=1$, electron interaction drives $t$TMD into a completely valley polarized state at zero external magnetic field, which would give rise to the QAH effect.

Experiments on $t$WSe$_2$ at twist angles $\theta \sim 5^\circ$ have found an interaction-induced insulating state at $n=1$~\cite{wang2020correlated,ghiotto2021quantum}. However, there is no spontaneous Hall effect at zero field, indicating the absence of valley polarization. This is likely due to the relatively large moir\'e bandwidth at the twist angles experimentally studied so far. To reduce the kinetic energy cost, an intervalley coherent state -- which nullifies the Berry curvature by hybridizing opposite valleys -- is found to be energetically favorable compared to the valley polarized state~\cite{bi2021excitonic,zang2021hartree,zang2022dynamical}.

A new opportunity opened up when recent theoretical modeling and large-scale density functional theory calculation~\cite{devakul2021magic} found that the topological bands in $t$WSe$_2$ become extremely flat near a ``magic-angle'' $\theta^* \simeq 1.45^\circ$. Such topological flat  band overcomes the previous limitation, and makes this system a promising candidate for realizing interaction-induced topological phases. Indeed, Hartree-Fock calculations found the QAH insulator at $n=1$ near this magic angle~\cite{devakul2021magic}, which arises from spontaneous complete spin/valley polarization.

In this work, we use a variety of numerical methods to study interaction-induced electronic phases in $t$TMD at $n<1$ fillings of the flat band. We find robust Ising ferromagnetism in the metallic state over a range of fillings and twist angles. The spin/valley polarization is driven by Coulomb repulsion in topological flat band. The completely polarized state extends up to relatively high temperature and to twist angle $\theta\sim 2.5^\circ$, a realistic regime for experiments.

We also show that a  spontaneous fractional Chern insulator, also known as fractional quantum anomalous Hall (FQAH) state, may be realized under suitable condition at $n=1/3$ filling of $t$WSe$_2$ near the magic angle. Importantly, the energy scale for FQAH at $n=1/3$ is much smaller than the spin gap in most of the filling factor range $n \leq 1$. Our work highlights that Ising ferromagnetism leading to anomalous Hall metal is a prerequisite to the highly sought-after FQAH state in $t$TMD, which was also recently predicted in twisted bilayer MoTe$_2$~\cite{li2021spontaneous}. Our work, however, shows the stringent requirements for its realization, namely the necessity of a magic angle and metallic gate to properly screen the Coulomb interaction which may otherwise favor a competing charge density wave.

\paragraph*{Generalized Kane-Mele model in $t$TMD ---}

\begin{figure}
\centering
\includegraphics[width=\columnwidth]{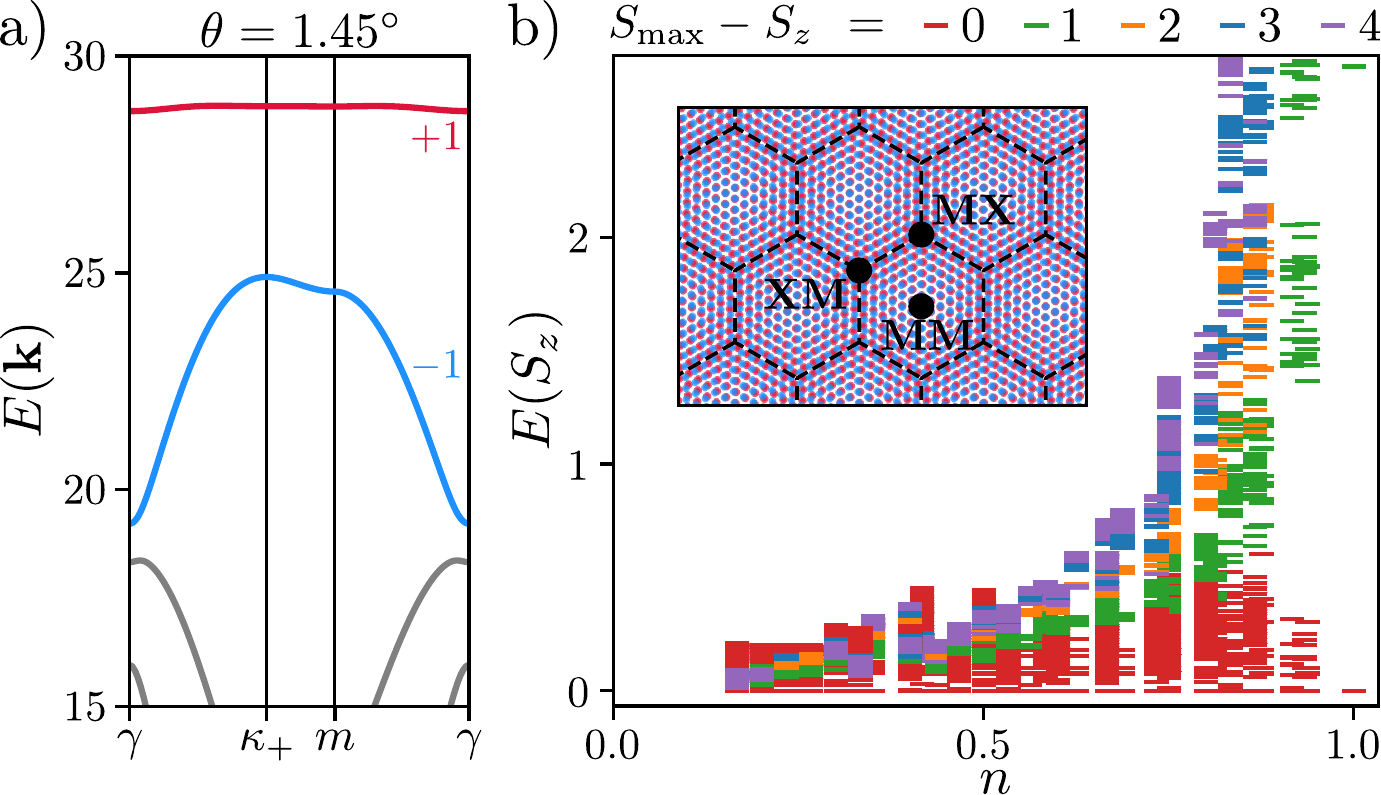}
\caption{a) $K$ valley moiré bands for a twisted WSe$_2$ homobilayer near the magic angle ($\theta = 1.45^\circ$), with colors indicating their Chern numbers. The $-K$ valleys bands are time reversal conjugate and carry opposite Chern number. b) All low-lying states of the interacting tight-binding model on the MX-XM honeycomb lattice (inset) are fully spin/valley polarized for densities $n \gtrsim 0.2$, as shown here for $U=\SI{25}{\milli\electronvolt} \gg w$. Only the lowest lying eigenstates obtained by ED are displayed in each sector, their color corresponding to the total spin. The results for several lattice size ($4\times3$, $4\times4$, $5\times3$, $5\times4$ and $6\times5$) with at least five particles are overlaid. Energies are given in meV, and measured with respect to the ferromagnetic ground state.
}
\label{fig:ContinuumModel}
\end{figure}

We study two WSe$_2$ layers twisted by a small angle $\theta$ with respect to the AA stacking. The resulting moiré pattern forms a honeycomb lattice with XM and MX sites, where the chalcogen atom X=Se on the top layer is aligned with the transition metal atom M=W on the bottom layer, or vice versa (see inset of Fig.~\ref{fig:ContinuumModel}). The XM and MX sites are related by a two-fold rotation symmetry interchanging the two layers.

The first and second moir\'e valence bands can thus be captured by an effective tight-binding model on the honeycomb lattice, originally introduced by Wu {\it et al.}~\cite{wu2019topological}. Since the Wannier orbital on the XM (MX) site resides primarily on the top (bottom) layer, the nearest-neighbor hopping between different sublattice sites is induced by interlayer tunneling, while second-nearest-neighbor hopping between same sublattice sites arises from the moir\'e potential within each layer. Due to the twist angle, the $\pm K$-valley band edges of the top and bottom layers are displaced from each other in ${\bf k}$-space and located at the two corners of the mini Brillouin zone $\kappa_\pm$. As a result, the second-neighbor hopping $t_2$ in the effective tight-binding model acquires a phase factor $e^{\pm i \phi}$ with $\phi \sim \vec{K} \cdot \vec{a}_M = 2\pi/3$ ($\vec{a}_M$ is a primitive moir\'e lattice vector), which is opposite for $\pm K$ valley. Due to spin/valley locking in the WSe$_2$ monolayers, the $\pm K$ valleys also correspond to opposite spins $s_z=\uparrow, \downarrow$.

The resulting model is therefore a generalization of the Kane-Mele model:
\begin{equation} \label{eq:LongrangeKaneMele}
\mathcal{H} = \sum_{n}  \sum_{\langle \vec{r}, \vec{r}' \rangle_n} (t_n c_{\vec{r},\uparrow}^\dagger c_{\vec{r}', \uparrow} + t_n^* c_{\vec{r}, \downarrow}^\dagger c_{\vec{r}', \downarrow} + hc) + V_n n_{\vec{r}} n_{\vec{r}'} .
\end{equation}
Here, $\langle \vec{r}, \vec{r}' \rangle_n$ denotes hopping between $n$-th neighbors on the honeycomb lattice. Importantly, long-range hoppings $t_{n=1,\cdots, 5}$ are included to faithfully reproduce the \textit{ab-initio} band structure determined by large-scale DFT~\cite{suppmat}. Amongst these tunneling amplitude, only $t_2$ has an appreciable imaginary component consistent with $\phi \sim 2\pi/3$. As a result, the first and second moiré valence bands are separated by a topological gap, and respectively carry a spin Chern number $C^1_s = 1$ and  $C^2_s = -1$ respectively, where $C_s \equiv C_\uparrow = -C_\downarrow$ follows from time-reversal symmetry (see Fig.~\ref{fig:ContinuumModel}a).

Of prime importance to our work is the $\theta$-dependent bandwidth of the topmost valence band~\cite{suppmat}, which exhibits a deep minimum near a ``magic-angle'' of $\theta^* \simeq 1.45^\circ$ where it reaches values $w \sim \SI{60}{\micro\electronvolt}$ (see Fig.~\ref{fig:ContinuumModel}a). The presence of  topological flat band makes $t$TMD susceptible to interaction-driven spin/valley polarization. In the following analysis, we shall first consider only  on-site repulsion, \textit{i.e.} $V_0 = U$ and $V_{n\geq1}=0$. This simplification captures the dominant part of the Coulomb potential for the small twist angles considered here, and can also be experimentally realized via gate screening, as shown in recent  experiments on TMD heterostructures~\cite{qiu2019giant,goodwin2019twist,kim2020control,xu2022tunable}.

\paragraph*{Half metal ---}

By exact diagonalization (ED) of Eq.~\ref{eq:LongrangeKaneMele} with on-site repulsion $U$, we find Ising ferromagnetism with complete spin/valley polarization in a wide range of fillings  $0.2<n \leq 1$ around the magic angle. The fully polarized ground state below unit filling is obtained by occupying the lowest energy orbitals of the topmost moir\'e band of one spin (=valley) component only, therefore avoiding $U$ completely. The resulting state is a ferromagnetic metal, also known as half metal. 

Fig.~\ref{fig:ContinuumModel}b shows the low-lying energy states at $\theta^*$, including several spin-flip ($S_{\rm max}-S_z = 0,\cdots,4$) above the fully spin-$\uparrow$ polarized sector. While we fix $U=\SI{25}{\milli\electronvolt}$, the precise value of the on-site repulsion does not matter as long as it largely exceeds all energy scales involved in the tight-biding model. Our results for various system sizes indicate a spin gap for $n\gtrsim 0.2$. Moreover, a manifold of fully polarized many-body states is present at energies below the spin gap, corresponding to particle-hole excitations around the Fermi surface.

The presence of a spin gap in the fully polarized metal (as opposed to gapless magnons) is made possible by the strong spin-orbit coupling in TMD, which reduces the $SU(2)$ spin symmetry to $U(1)$. Our ED study shows that the lowest energy spin excitation corresponds to a single spin flip, hence the spin gap is given by $\Delta_{\rm SG} = E(S_{\rm max}-1) - E(S_{\rm max})$. 
A physical picture for the spin gap in $t$WSe$_2$ is that a spin-$\downarrow$ fermion cannot completely avoid overlap with all spin-$\uparrow$ fermions and completely avoid kinetic energy cost.

\begin{figure}
\centering
\includegraphics[width=\columnwidth]{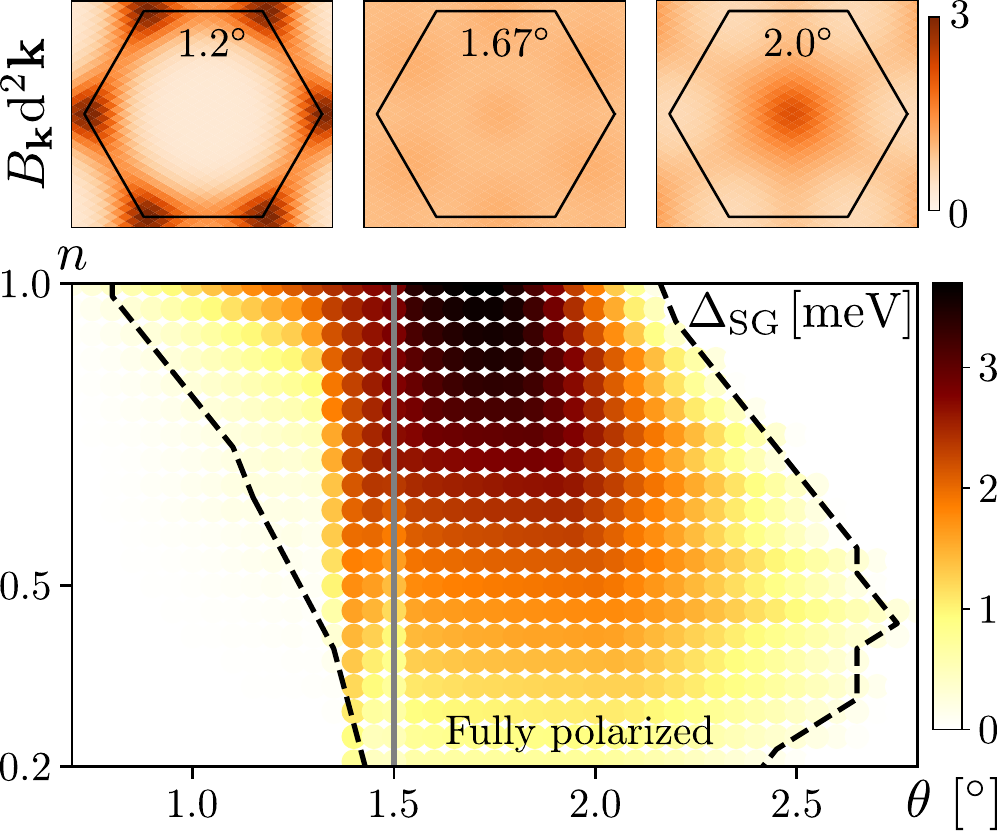}
\caption{Spin-gap $\Delta_{\rm SP}$ obtained from a spin-flip analysis on the tight-binding ($\theta<1.5^\circ$) and continuum ($\theta>1.5^\circ$) models with 576 point in the moiré Brillouin zone. The dashed line encircles the region in which the system is fully spin/valley polarized. The spin gap is largest where the Berry curvature's fluctuations are minimal (insets), showing a close connection between ferromagnetism and topology. 
}
\label{fig:FMMphase}
\end{figure}

A necessary condition for spontaneous and complete spin/valley polarization is $\Delta_{\rm SG} > 0$. By calculating $\Delta_{\rm SG}$ with a spin-flip analysis (see supplementary materials~\cite{suppmat}), we identify the region of half metal as a function of filling factor and twist angle $\theta < 1.5^\circ$, shown in Fig.~\ref{fig:FMMphase}. Large values of the gap indicate Ising ferromagnetism persisting at higher temperatures, which is found near unit filling $n=1$ and close to the magic angle $\theta^*$.

For $\theta > 1.5^\circ$, the first and second moiré valence bands carry identical spin Chern numbers $C^1_s=C_s^2$, which cannot be captured by the two-band model Eq.~\ref{eq:LongrangeKaneMele}. To extend our analysis to larger twist angles, we study the continuum model of WSe$_2$ homobilayers~\cite{wu2019topological,devakul2021magic} with electron interaction. The spin-flip analysis can be applied to the continuum if the gate-screened Coulomb interaction is approximated by a contact repulsion~\cite{alavirad2020ferromagnetism}. Despite being a crude approximation, it nevertheless reproduces essential features of our lattice model result~\cite{suppmat}, and can therefore be used as a qualitative guide to the half metal.

The spin-flip result for the interacting continuum model are shown in Fig.~\ref{fig:FMMphase} for $\theta>1.5^\circ$, where we observe that the half metal extends up to $\theta \sim 2.5^\circ$ and in a wide range of densities. The spin gap reaches maximal values $\sim 3.5$meV, which suggests a ferromagnetic critical temperature $T_{\rm FM}$ on the order of $\SI{40}{\kelvin}$. We also observe that the spin gap is maximum for densities below unit filling when $\theta>1.75^\circ$, indicating that the half metal can be more robust than the QAH insulator at $n=1$. We finally notice that the spin gap qualitatively correlates with the variance of the Berry curvature (insets of Fig.~\ref{fig:FMMphase}), suggesting an intimate relation between topology and Ising ferromagnetism in our model.

To summarize the first part of this work, we predict a robust half metallic phase in twisted WSe$_2$ in the experimentally accessible regime $\theta < 2.5^\circ$ and $n<1$. This phase is stabilized by the short-ranged and dominant part of the Coulomb repulsion, which enables a relatively high ferromagnetic temperature that can reach a few tens of Kelvins. Thanks to the spin-valley locking in TMD, the Ising ferromagnetism can be detected optically through circular dichroism. The presence of Berry curvature and band topology is manifested by the anomalous Hall effect in the half metal.

The on-site repulsion $U$ is crucial in inducing complete spin/valley polarization with a spin gap, but has no effects in the fully spin/valley polarized phase. As a result, below $T_{\rm FM}$, the subdominant components of the interactions $V_{n\geq 1}$, up to now neglected, may have important effect in determining the ground state of the system. If $V_{n\geq 1}$ are small compared to the spin gap, the low temperature phase will remain fully polarized. Therefore, thanks to the hierarchy in energy scales between the short and long range part of the Coulomb repulsion on the moiré lattice, the half metal that onsets at high temperature is a precursor to and a parent state of low-temperature phases in $t$WSe$_2$, which we now turn to.

\paragraph*{Fractional quantum anomalous Hall insulator ---}

We now show by comprehensive ED study that a spontaneous fractional Chern insulator also known as fractional quantum anomalous Hall (FQAH) state may arise in twisted WSe$_2$ homobilayer. Its appearance follows the general argument given above: (i) The system at fractional filling $n=1/3$ shows spontaneous complete spin/valley polarization around $\theta^*$ with a spin-gap $\SI{0.13}{\milli\electronvolt}$ (see Fig.~\ref{fig:FMMphase}); (ii) In the fully polarized phase, the generalized Kane-Mele model Eq.~\ref{eq:LongrangeKaneMele} reduces to a Haldane model with long range hoppings~\cite{haldane1988model}, for which consistent evidence from ED~\cite{wu2012zoology,dobardvzic2013geometrical,bernevig2012emergent} and density-matrix renormalization group (DMRG)~\cite{grushin2015characterization} have observed the emergence of an fractional chern insulator.

To put these ideas on firm grounds, we first consider $V_1 = V = \SI{3.5}{\milli\electronvolt}$ and $V_{n>1}=0$ keeping $U=\SI{25}{\milli\electronvolt}$ as before, and perform extensive ED calculations to find the ground state of Eq.~\ref{eq:LongrangeKaneMele} projected to the first moiré band~\cite{regnault2011fractional,sheng2011fractional,neupert2011fractional,crepel2020microscopic}. We perform calculations on finite clusters with $N_1 \times N_2$ unit cells along the two moiré lattice basis vectors, with a total number of holes equal to $N = N_\uparrow + N_\downarrow = N_1 N_2 / 3$. Translation invariance allows to resolve the many-body momenta $(K_1, K_2)$ along the two moiré basis vectors.

\begin{figure}
\centering
\includegraphics[width=\columnwidth]{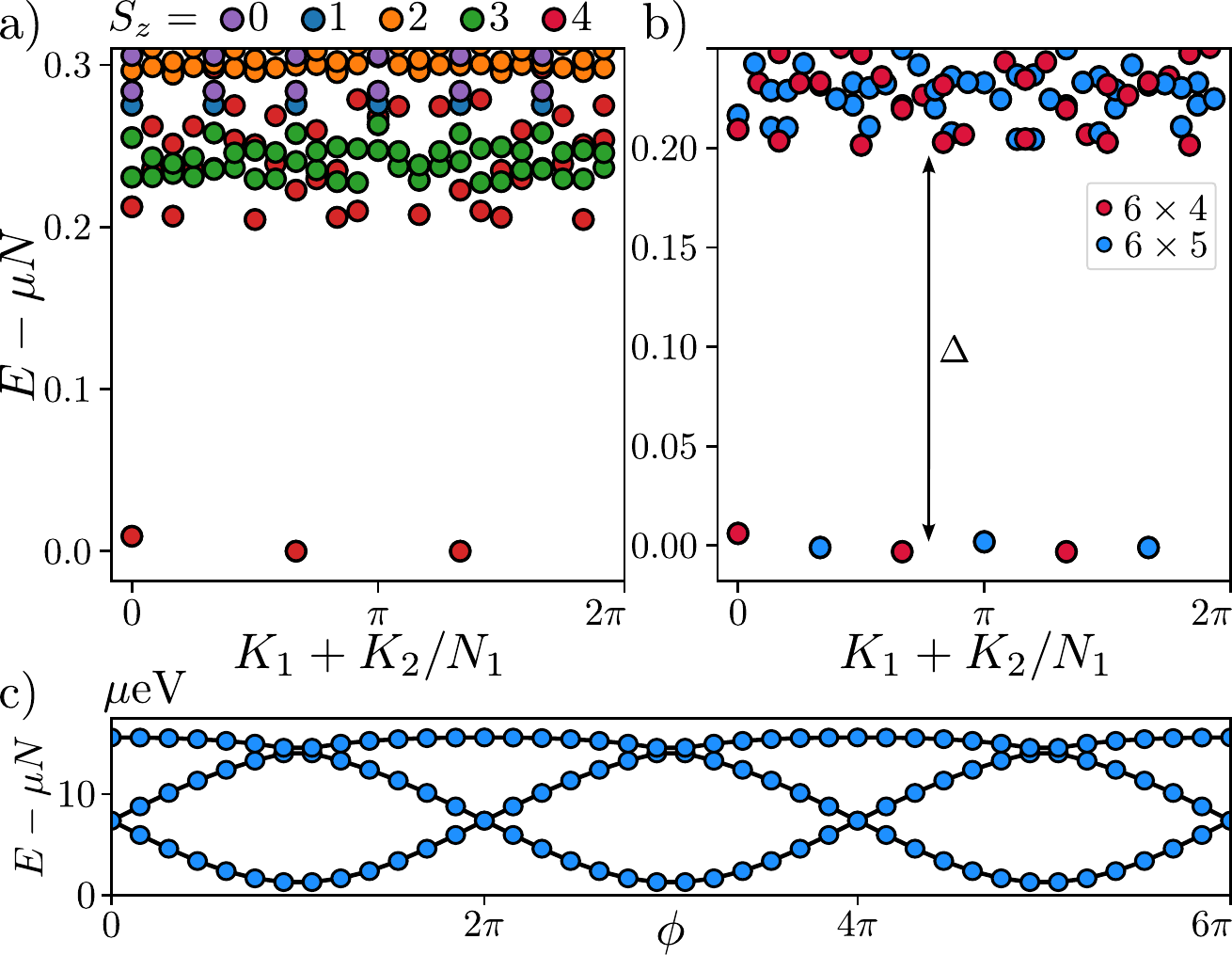}
\caption{
a) Many-body spectrum at filling $n=1/3$ as a function of the many-body momentum $(K_1, K_2)$ for $V = \SI{3.5}{\milli\electronvolt}$ and $\theta=1.4^\circ$, obtained with ED on $6 \times 4$ lattice including all spin sectors. energies are given in meV. The system remains fully spin/valley polarized. b) Same as (a) in the fully spin-polarized sector and for larger system sizes. Three low-lying states clearly detach from the continuum. c) Their spectral flow exhibits the characteristic $6\pi$-periodicity of FCI ground states.
}
\label{fig:FirstHintFCI}
\end{figure}

For consistency, we first check that the system is fully spin/valley polarized. Our ED results at the magic-angle on the $6\times4$ lattice including all spin sectors, shown in Fig.~\ref{fig:FirstHintFCI}a, prove that the model at filling $n=1/3$ is indeed ferromagnetic. The energy of a single spin-flip  extracted from these data $\Delta_{\rm SG} \simeq \SI{0.15}{\milli\electronvolt}$ almost quantitatively match the value obtained in Fig.~\ref{fig:FMMphase} for $n=1/3$ at $V=0$ and $\theta = 1.45^\circ$.

We also notice three low-lying and almost degenerate states, which are clearly detached from the many-body continuum and appear for all considered system sizes $6\times N_2$ with $N_2=4,5,6$ (Fig.~\ref{fig:FirstHintFCI}b). They are located at many-body momenta $K_1 = \frac{2\pi}{3} [0,k,2k] + Q_1$ and $K_2 = 0$, with a Jordan-Wigner shift $Q_1 = 0$ if $N_2$ is even and $Q_1 = \pi/6$ if it is odd, which precisely match the counting rule of a Laughlin-like FCI~\cite{regnault2011fractional,bernevig2012emergent,crepel2018matrix}. As another evidence of the FCI behavior, we compute the anomalous Hall conductance associated with the ground state manifold  $\sigma = \frac{e^2}{h} \frac{2\pi}{N_1 N_2} \sum_{\vec{k}} n_{\vec{k}} B_{\vec{k}}$, with $n_{\vec{k}}$ the mean occupation of the $\vec{k}$ single-particle state in the three nearly-degenerate ground states, and $B_{\vec{k}}$ the single particle Berry curvature~\cite{grushin2012enhancing,neupert2012elementary}. In units of $e^2 C_s^1/h$, we find $\sigma_{xy}= 0.33$, $0.32$ and $0.31$ for $N_2\in\{4,5,6\}$~\footnote{In this calculation, both the Berry curvature $B_{\vec{k}}$ and the Chern number $C_\uparrow$ were computed with the finite momentum discretization imposed by the $6 \times N_2$ lattice, using the method of Ref.~\cite{fukui2005chern}.}, in excellent agreement with the theoretical expectation of $\sigma_{xy}^{\rm FCI}=1/3$.

Finally, a necessary condition for the emergence of the FCI -- which is also sufficient if the gap $\Delta$ does not close in the thermodynamic limit~\cite{regnault2011fractional,crepel2019matrix}\footnote{While the gap $\Delta$ appears robust in our calculations, a finite size scaling of the gap to the thermodynamics is however plagued by commensuration effects stabilizing CDW when both $N_1$ and $N_2$ are multiples of six (Fig.~\ref{fig:FirstHintFCI}b). See Refs.~\cite{wu2012zoology} and~\cite{grushin2015characterization} for a longer discussion.} -- is that each topological ground state must undergo cyclic permutation when $\phi$ goes from $0$ to $2\pi$ such that it comes back only after three units of flux ($\phi = 6\pi$)~\cite{thouless1989level}. As shown in Fig.~\ref{fig:FirstHintFCI}c for the $6\times5$ lattice, this necessary condition is satisfied in our model. Altogether, the numerical evidence presented in Fig.~\ref{fig:FirstHintFCI}a-c clearly identifies an FQAH phase at fractional filling of our $t$TMD, with a spontaneous spin/valley polarization inherited from the parent half metal. A similar FQAH state was found in $t$MoTe$_2$ using a different numerical method \cite{li2021spontaneous}.

\begin{figure}
\centering
\includegraphics[width=\columnwidth]{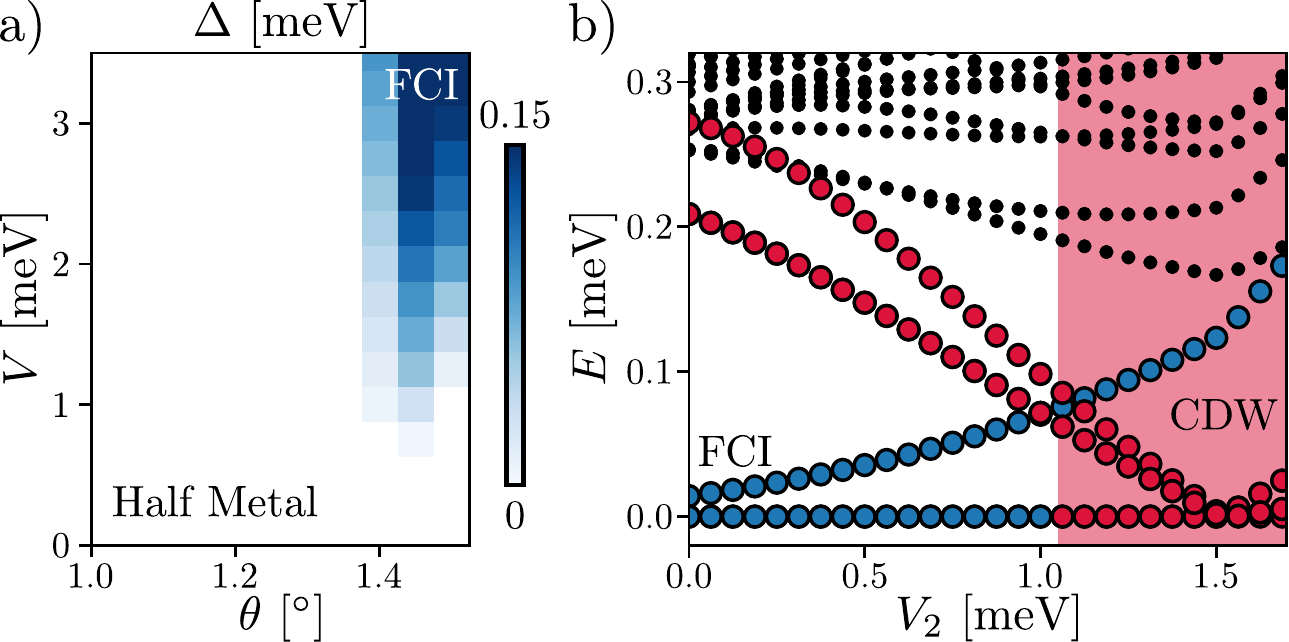}
\caption{a) FQAH many-body gap as a function of interaction strength $V$ and twist angle $\theta$. The half metal dominate the phase diagram except near the magic angle where FCI behaviors emerge. b) Longer range interaction drive a transition from FQAH to CDW. For visibility, we have colored the state connected to the FCI (resp. CDW) in blue (resp. red). 
}
\label{fig:NeighboringPhases}
\end{figure}

To further probe the robustness of the FQAH phase, we compute the many-body gap $\Delta$ above its ground state manifold (defined in Fig.~\ref{fig:FirstHintFCI}) as a function interaction strength and twist angle. Our results presented in Fig.~\ref{fig:NeighboringPhases}a are consistent with our previous findings; near the magic-angle, the half metal found at $V=0$ (Fig.~\ref{fig:FMMphase}) becomes unstable to a FCI when $V$ increases (Fig.~\ref{fig:FirstHintFCI}). The situation quickly becomes unfavorable for the FQAH as one moves away from the magic angle, with $\Delta$ almost halved when $\theta$ goes away from the magic-angle by $\delta\theta = 0.05^\circ$. Even worse, for deviations $\theta \sim 0.15^\circ$, the complete spin/valley ferromagnetism is lost as the spin gap at $n=1/3$ closes near $\theta = 1.3^\circ$ (Fig.~\ref{fig:ContinuumModel}). The observation of a FQAH in $t$WSe$_2$ thus requires stringent experimental conditions, and in particular accurate control of the twist angle.

Another perturbation reducing the FQAH many-body gap comes from the longer range part of the interaction $V_{n>1}$. Indeed, the ground state at filling $n=1/3$ when $V_1=V_2=\infty$ is a charge density wave (CDW) of $\sqrt{3} \times \sqrt{3}$ periodicity~\cite{zhang2021electronic}. A FQAH to CDW transition is thus expected as $V_2$ increases. Because such CDW only occupies one of the honeycomb sublattice, band projected ED cannot faithfully describe it as the topmost valence band is sublattice hybridized. To resolve this issue, we perform ED in real-space using a 6$\times$4 lattice. To keep the Hilbert space manageable, we discard any configuration with a non-zero number of double occupation or nearest neighbor pair ~\cite{kourtis2014fractional,grushin2015characterization}.

The results of this analysis, depicted in Fig.~\ref{fig:NeighboringPhases}b, show a transition from a three-fold to a six-fold degenerate ground state as $V_2$ increases~\footnote{The inversion symmetry of the model makes states at momenta $(K_1,K_2)$ and $(-K_1,-K_2)$ coincide.}. The former satisfies the Laughlin-like counting rule and represents the previously identified FQAH, while the degeneracy of the latter perfectly match the expectation for a CDW on the honeycomb lattice. We find that the transition between FQAH and CDW occurs around $V_2^c \simeq \SI{1.05}{\milli\electronvolt}$. For $V_1 = \SI{3.5}{\milli\electronvolt}$, this corresponds to  $V_1 \simeq 1.8 \sqrt{3} V_2^c$. Therefore, the realization of the FQAH requires screening from nearby metallic gates to decrease the $V_1/V_2$ ratio by a factor of at least two compared to its Coulomb value. The competition between FQAH and CDW states and the importance of gate screening have been overlooked in the previous study on $t$TMD.    

Our FQAH state with spontaneous spin/valley polarization at zero field should be contrasted with the FCIs reported in graphene/hBN and magic-angle twisted bilayer graphene, which rely on a sufficiently large external magnetic fields. For instance, the most recent experiment on magic-angle graphene~\cite{xie2021fractional} finds an incompressible state with \textit{zero} Hall conductance for magnetic fields below $\SI{5}{\tesla}$ at $3+1/3$ filling of the flat band, while the evidence for FCIs---fractional slope and fractional intercept of incompressible states---shows up at higher field.  As an alternative scenario however, the observed feature could also arise if a CDW at $1/3$ filling of a trivial flat band coexists with a $\nu=1/3$ fractional quantum Hall state from Landau levels of a dispersive band. The search for anomalous Hall effect~\cite{xie2021weak} and  FQAH state in twisted bilayer graphene at fractional fillings~\cite{ledwith2020fractional,repellin2020chern,liu2021gate} continues.

Our work highlights the importance of anomalous Hall metal state as a precursor to FQAH in TMD moir\'e heterostructures having topological flat bands. While our calculations were performed for $t$WSe$_2$, our main conclusion also applies to other twisted TMD homobilayers, although the magic angle and the maximum spin gap are material dependent.

\paragraph*{Acknowledgement ---}  We are grateful to Trithep Devakul and Yang Zhang for a previous collaboration that motivated and paved the way for this study. Our work is supported by the Simons Foundation through a Simons Investigator Award. L.F. is partly supported by the David and Lucile Packard Foundation. V.C. gratefully acknowledges support from the MathWorks fellowship.

\bibliography{BiblioFCI}

\begin{thebibliography}{48}%
\makeatletter
\providecommand \@ifxundefined [1]{%
 \@ifx{#1\undefined}
}%
\providecommand \@ifnum [1]{%
 \ifnum #1\expandafter \@firstoftwo
 \else \expandafter \@secondoftwo
 \fi
}%
\providecommand \@ifx [1]{%
 \ifx #1\expandafter \@firstoftwo
 \else \expandafter \@secondoftwo
 \fi
}%
\providecommand \natexlab [1]{#1}%
\providecommand \enquote  [1]{``#1''}%
\providecommand \bibnamefont  [1]{#1}%
\providecommand \bibfnamefont [1]{#1}%
\providecommand \citenamefont [1]{#1}%
\providecommand \href@noop [0]{\@secondoftwo}%
\providecommand \href [0]{\begingroup \@sanitize@url \@href}%
\providecommand \@href[1]{\@@startlink{#1}\@@href}%
\providecommand \@@href[1]{\endgroup#1\@@endlink}%
\providecommand \@sanitize@url [0]{\catcode `\\12\catcode `\$12\catcode
  `\&12\catcode `\#12\catcode `\^12\catcode `\_12\catcode `\%12\relax}%
\providecommand \@@startlink[1]{}%
\providecommand \@@endlink[0]{}%
\providecommand \url  [0]{\begingroup\@sanitize@url \@url }%
\providecommand \@url [1]{\endgroup\@href {#1}{\urlprefix }}%
\providecommand \urlprefix  [0]{URL }%
\providecommand \Eprint [0]{\href }%
\providecommand \doibase [0]{https://doi.org/}%
\providecommand \selectlanguage [0]{\@gobble}%
\providecommand \bibinfo  [0]{\@secondoftwo}%
\providecommand \bibfield  [0]{\@secondoftwo}%
\providecommand \translation [1]{[#1]}%
\providecommand \BibitemOpen [0]{}%
\providecommand \bibitemStop [0]{}%
\providecommand \bibitemNoStop [0]{.\EOS\space}%
\providecommand \EOS [0]{\spacefactor3000\relax}%
\providecommand \BibitemShut  [1]{\csname bibitem#1\endcsname}%
\let\auto@bib@innerbib\@empty
\bibitem [{\citenamefont {Wu}\ \emph {et~al.}(2018)\citenamefont {Wu},
  \citenamefont {Lovorn}, \citenamefont {Tutuc},\ and\ \citenamefont
  {MacDonald}}]{wu2018hubbard}%
  \BibitemOpen
  \bibfield  {author} {\bibinfo {author} {\bibfnamefont {F.}~\bibnamefont
  {Wu}}, \bibinfo {author} {\bibfnamefont {T.}~\bibnamefont {Lovorn}}, \bibinfo
  {author} {\bibfnamefont {E.}~\bibnamefont {Tutuc}},\ and\ \bibinfo {author}
  {\bibfnamefont {A.~H.}\ \bibnamefont {MacDonald}},\ }\href@noop {} {\bibfield
   {journal} {\bibinfo  {journal} {Physical review letters}\ }\textbf {\bibinfo
  {volume} {121}},\ \bibinfo {pages} {026402} (\bibinfo {year}
  {2018})}\BibitemShut {NoStop}%
\bibitem [{\citenamefont {Zhang}\ \emph {et~al.}(2020)\citenamefont {Zhang},
  \citenamefont {Yuan},\ and\ \citenamefont {Fu}}]{zhang2020moire}%
  \BibitemOpen
  \bibfield  {author} {\bibinfo {author} {\bibfnamefont {Y.}~\bibnamefont
  {Zhang}}, \bibinfo {author} {\bibfnamefont {N.~F.}\ \bibnamefont {Yuan}},\
  and\ \bibinfo {author} {\bibfnamefont {L.}~\bibnamefont {Fu}},\ }\href@noop
  {} {\bibfield  {journal} {\bibinfo  {journal} {Physical Review B}\ }\textbf
  {\bibinfo {volume} {102}},\ \bibinfo {pages} {201115} (\bibinfo {year}
  {2020})}\BibitemShut {NoStop}%
\bibitem [{\citenamefont {Tang}\ \emph {et~al.}(2020)\citenamefont {Tang},
  \citenamefont {Li}, \citenamefont {Li}, \citenamefont {Xu}, \citenamefont
  {Liu}, \citenamefont {Barmak}, \citenamefont {Watanabe}, \citenamefont
  {Taniguchi}, \citenamefont {MacDonald}, \citenamefont {Shan} \emph
  {et~al.}}]{tang2020simulation}%
  \BibitemOpen
  \bibfield  {author} {\bibinfo {author} {\bibfnamefont {Y.}~\bibnamefont
  {Tang}}, \bibinfo {author} {\bibfnamefont {L.}~\bibnamefont {Li}}, \bibinfo
  {author} {\bibfnamefont {T.}~\bibnamefont {Li}}, \bibinfo {author}
  {\bibfnamefont {Y.}~\bibnamefont {Xu}}, \bibinfo {author} {\bibfnamefont
  {S.}~\bibnamefont {Liu}}, \bibinfo {author} {\bibfnamefont {K.}~\bibnamefont
  {Barmak}}, \bibinfo {author} {\bibfnamefont {K.}~\bibnamefont {Watanabe}},
  \bibinfo {author} {\bibfnamefont {T.}~\bibnamefont {Taniguchi}}, \bibinfo
  {author} {\bibfnamefont {A.~H.}\ \bibnamefont {MacDonald}}, \bibinfo {author}
  {\bibfnamefont {J.}~\bibnamefont {Shan}}, \emph {et~al.},\ }\href@noop {}
  {\bibfield  {journal} {\bibinfo  {journal} {Nature}\ }\textbf {\bibinfo
  {volume} {579}},\ \bibinfo {pages} {353} (\bibinfo {year}
  {2020})}\BibitemShut {NoStop}%
\bibitem [{\citenamefont {Regan}\ \emph {et~al.}(2020)\citenamefont {Regan},
  \citenamefont {Wang}, \citenamefont {Jin}, \citenamefont {Utama},
  \citenamefont {Gao}, \citenamefont {Wei}, \citenamefont {Zhao}, \citenamefont
  {Zhao}, \citenamefont {Zhang}, \citenamefont {Yumigeta} \emph
  {et~al.}}]{regan2020mott}%
  \BibitemOpen
  \bibfield  {author} {\bibinfo {author} {\bibfnamefont {E.~C.}\ \bibnamefont
  {Regan}}, \bibinfo {author} {\bibfnamefont {D.}~\bibnamefont {Wang}},
  \bibinfo {author} {\bibfnamefont {C.}~\bibnamefont {Jin}}, \bibinfo {author}
  {\bibfnamefont {M.~I.~B.}\ \bibnamefont {Utama}}, \bibinfo {author}
  {\bibfnamefont {B.}~\bibnamefont {Gao}}, \bibinfo {author} {\bibfnamefont
  {X.}~\bibnamefont {Wei}}, \bibinfo {author} {\bibfnamefont {S.}~\bibnamefont
  {Zhao}}, \bibinfo {author} {\bibfnamefont {W.}~\bibnamefont {Zhao}}, \bibinfo
  {author} {\bibfnamefont {Z.}~\bibnamefont {Zhang}}, \bibinfo {author}
  {\bibfnamefont {K.}~\bibnamefont {Yumigeta}}, \emph {et~al.},\ }\href@noop {}
  {\bibfield  {journal} {\bibinfo  {journal} {Nature}\ }\textbf {\bibinfo
  {volume} {579}},\ \bibinfo {pages} {359} (\bibinfo {year}
  {2020})}\BibitemShut {NoStop}%
\bibitem [{\citenamefont {Xu}\ \emph {et~al.}(2020)\citenamefont {Xu},
  \citenamefont {Liu}, \citenamefont {Rhodes}, \citenamefont {Watanabe},
  \citenamefont {Taniguchi}, \citenamefont {Hone}, \citenamefont {Elser},
  \citenamefont {Mak},\ and\ \citenamefont {Shan}}]{xu2020correlated}%
  \BibitemOpen
  \bibfield  {author} {\bibinfo {author} {\bibfnamefont {Y.}~\bibnamefont
  {Xu}}, \bibinfo {author} {\bibfnamefont {S.}~\bibnamefont {Liu}}, \bibinfo
  {author} {\bibfnamefont {D.~A.}\ \bibnamefont {Rhodes}}, \bibinfo {author}
  {\bibfnamefont {K.}~\bibnamefont {Watanabe}}, \bibinfo {author}
  {\bibfnamefont {T.}~\bibnamefont {Taniguchi}}, \bibinfo {author}
  {\bibfnamefont {J.}~\bibnamefont {Hone}}, \bibinfo {author} {\bibfnamefont
  {V.}~\bibnamefont {Elser}}, \bibinfo {author} {\bibfnamefont {K.~F.}\
  \bibnamefont {Mak}},\ and\ \bibinfo {author} {\bibfnamefont {J.}~\bibnamefont
  {Shan}},\ }\href@noop {} {\bibfield  {journal} {\bibinfo  {journal} {Nature}\
  }\textbf {\bibinfo {volume} {587}},\ \bibinfo {pages} {214} (\bibinfo {year}
  {2020})}\BibitemShut {NoStop}%
\bibitem [{\citenamefont {Huang}\ \emph {et~al.}(2021)\citenamefont {Huang},
  \citenamefont {Wang}, \citenamefont {Miao}, \citenamefont {Wang},
  \citenamefont {Li}, \citenamefont {Lian}, \citenamefont {Taniguchi},
  \citenamefont {Watanabe}, \citenamefont {Okamoto}, \citenamefont {Xiao} \emph
  {et~al.}}]{huang2021correlated}%
  \BibitemOpen
  \bibfield  {author} {\bibinfo {author} {\bibfnamefont {X.}~\bibnamefont
  {Huang}}, \bibinfo {author} {\bibfnamefont {T.}~\bibnamefont {Wang}},
  \bibinfo {author} {\bibfnamefont {S.}~\bibnamefont {Miao}}, \bibinfo {author}
  {\bibfnamefont {C.}~\bibnamefont {Wang}}, \bibinfo {author} {\bibfnamefont
  {Z.}~\bibnamefont {Li}}, \bibinfo {author} {\bibfnamefont {Z.}~\bibnamefont
  {Lian}}, \bibinfo {author} {\bibfnamefont {T.}~\bibnamefont {Taniguchi}},
  \bibinfo {author} {\bibfnamefont {K.}~\bibnamefont {Watanabe}}, \bibinfo
  {author} {\bibfnamefont {S.}~\bibnamefont {Okamoto}}, \bibinfo {author}
  {\bibfnamefont {D.}~\bibnamefont {Xiao}}, \emph {et~al.},\ }\href@noop {}
  {\bibfield  {journal} {\bibinfo  {journal} {Nature Physics}\ }\textbf
  {\bibinfo {volume} {17}},\ \bibinfo {pages} {715} (\bibinfo {year}
  {2021})}\BibitemShut {NoStop}%
\bibitem [{\citenamefont {Li}\ \emph {et~al.}(2021{\natexlab{a}})\citenamefont
  {Li}, \citenamefont {Li}, \citenamefont {Regan}, \citenamefont {Wang},
  \citenamefont {Zhao}, \citenamefont {Kahn}, \citenamefont {Yumigeta},
  \citenamefont {Blei}, \citenamefont {Taniguchi}, \citenamefont {Watanabe}
  \emph {et~al.}}]{li2021imaging}%
  \BibitemOpen
  \bibfield  {author} {\bibinfo {author} {\bibfnamefont {H.}~\bibnamefont
  {Li}}, \bibinfo {author} {\bibfnamefont {S.}~\bibnamefont {Li}}, \bibinfo
  {author} {\bibfnamefont {E.~C.}\ \bibnamefont {Regan}}, \bibinfo {author}
  {\bibfnamefont {D.}~\bibnamefont {Wang}}, \bibinfo {author} {\bibfnamefont
  {W.}~\bibnamefont {Zhao}}, \bibinfo {author} {\bibfnamefont {S.}~\bibnamefont
  {Kahn}}, \bibinfo {author} {\bibfnamefont {K.}~\bibnamefont {Yumigeta}},
  \bibinfo {author} {\bibfnamefont {M.}~\bibnamefont {Blei}}, \bibinfo {author}
  {\bibfnamefont {T.}~\bibnamefont {Taniguchi}}, \bibinfo {author}
  {\bibfnamefont {K.}~\bibnamefont {Watanabe}}, \emph {et~al.},\ }\href@noop {}
  {\bibfield  {journal} {\bibinfo  {journal} {Nature}\ }\textbf {\bibinfo
  {volume} {597}},\ \bibinfo {pages} {650} (\bibinfo {year}
  {2021}{\natexlab{a}})}\BibitemShut {NoStop}%
\bibitem [{\citenamefont {Li}\ \emph {et~al.}(2021{\natexlab{b}})\citenamefont
  {Li}, \citenamefont {Jiang}, \citenamefont {Shen}, \citenamefont {Zhang},
  \citenamefont {Li}, \citenamefont {Tao}, \citenamefont {Devakul},
  \citenamefont {Watanabe}, \citenamefont {Taniguchi}, \citenamefont {Fu} \emph
  {et~al.}}]{li2021quantum}%
  \BibitemOpen
  \bibfield  {author} {\bibinfo {author} {\bibfnamefont {T.}~\bibnamefont
  {Li}}, \bibinfo {author} {\bibfnamefont {S.}~\bibnamefont {Jiang}}, \bibinfo
  {author} {\bibfnamefont {B.}~\bibnamefont {Shen}}, \bibinfo {author}
  {\bibfnamefont {Y.}~\bibnamefont {Zhang}}, \bibinfo {author} {\bibfnamefont
  {L.}~\bibnamefont {Li}}, \bibinfo {author} {\bibfnamefont {Z.}~\bibnamefont
  {Tao}}, \bibinfo {author} {\bibfnamefont {T.}~\bibnamefont {Devakul}},
  \bibinfo {author} {\bibfnamefont {K.}~\bibnamefont {Watanabe}}, \bibinfo
  {author} {\bibfnamefont {T.}~\bibnamefont {Taniguchi}}, \bibinfo {author}
  {\bibfnamefont {L.}~\bibnamefont {Fu}}, \emph {et~al.},\ }\href@noop {}
  {\bibfield  {journal} {\bibinfo  {journal} {Nature}\ }\textbf {\bibinfo
  {volume} {600}},\ \bibinfo {pages} {641} (\bibinfo {year}
  {2021}{\natexlab{b}})}\BibitemShut {NoStop}%
\bibitem [{\citenamefont {Wu}\ \emph {et~al.}(2019)\citenamefont {Wu},
  \citenamefont {Lovorn}, \citenamefont {Tutuc}, \citenamefont {Martin},\ and\
  \citenamefont {MacDonald}}]{wu2019topological}%
  \BibitemOpen
  \bibfield  {author} {\bibinfo {author} {\bibfnamefont {F.}~\bibnamefont
  {Wu}}, \bibinfo {author} {\bibfnamefont {T.}~\bibnamefont {Lovorn}}, \bibinfo
  {author} {\bibfnamefont {E.}~\bibnamefont {Tutuc}}, \bibinfo {author}
  {\bibfnamefont {I.}~\bibnamefont {Martin}},\ and\ \bibinfo {author}
  {\bibfnamefont {A.}~\bibnamefont {MacDonald}},\ }\href@noop {} {\bibfield
  {journal} {\bibinfo  {journal} {Physical review letters}\ }\textbf {\bibinfo
  {volume} {122}},\ \bibinfo {pages} {086402} (\bibinfo {year}
  {2019})}\BibitemShut {NoStop}%
\bibitem [{\citenamefont {Kane}\ and\ \citenamefont
  {Mele}(2005)}]{kane2005quantum}%
  \BibitemOpen
  \bibfield  {author} {\bibinfo {author} {\bibfnamefont {C.~L.}\ \bibnamefont
  {Kane}}\ and\ \bibinfo {author} {\bibfnamefont {E.~J.}\ \bibnamefont
  {Mele}},\ }\href@noop {} {\bibfield  {journal} {\bibinfo  {journal} {Physical
  review letters}\ }\textbf {\bibinfo {volume} {95}},\ \bibinfo {pages}
  {226801} (\bibinfo {year} {2005})}\BibitemShut {NoStop}%
\bibitem [{\citenamefont {Wang}\ \emph {et~al.}(2020)\citenamefont {Wang},
  \citenamefont {Shih}, \citenamefont {Ghiotto}, \citenamefont {Xian},
  \citenamefont {Rhodes}, \citenamefont {Tan}, \citenamefont {Claassen},
  \citenamefont {Kennes}, \citenamefont {Bai}, \citenamefont {Kim},
  \citenamefont {Watanabe}, \citenamefont {Taniguchi}, \citenamefont {Zhu},
  \citenamefont {Hone}, \citenamefont {Rubio}, \citenamefont {Pasupathy},\ and\
  \citenamefont {Dean}}]{wang2020correlated}%
  \BibitemOpen
  \bibfield  {author} {\bibinfo {author} {\bibfnamefont {L.}~\bibnamefont
  {Wang}}, \bibinfo {author} {\bibfnamefont {E.-M.}\ \bibnamefont {Shih}},
  \bibinfo {author} {\bibfnamefont {A.}~\bibnamefont {Ghiotto}}, \bibinfo
  {author} {\bibfnamefont {L.}~\bibnamefont {Xian}}, \bibinfo {author}
  {\bibfnamefont {D.~A.}\ \bibnamefont {Rhodes}}, \bibinfo {author}
  {\bibfnamefont {C.}~\bibnamefont {Tan}}, \bibinfo {author} {\bibfnamefont
  {M.}~\bibnamefont {Claassen}}, \bibinfo {author} {\bibfnamefont {D.~M.}\
  \bibnamefont {Kennes}}, \bibinfo {author} {\bibfnamefont {Y.}~\bibnamefont
  {Bai}}, \bibinfo {author} {\bibfnamefont {B.}~\bibnamefont {Kim}}, \bibinfo
  {author} {\bibfnamefont {K.}~\bibnamefont {Watanabe}}, \bibinfo {author}
  {\bibfnamefont {T.}~\bibnamefont {Taniguchi}}, \bibinfo {author}
  {\bibfnamefont {X.}~\bibnamefont {Zhu}}, \bibinfo {author} {\bibfnamefont
  {J.}~\bibnamefont {Hone}}, \bibinfo {author} {\bibfnamefont {A.}~\bibnamefont
  {Rubio}}, \bibinfo {author} {\bibfnamefont {A.~N.}\ \bibnamefont
  {Pasupathy}},\ and\ \bibinfo {author} {\bibfnamefont {C.~R.}\ \bibnamefont
  {Dean}},\ }\href {https://doi.org/10.1038/s41563-020-0708-6} {\bibfield
  {journal} {\bibinfo  {journal} {Nature Materials}\ }\textbf {\bibinfo
  {volume} {19}},\ \bibinfo {pages} {861} (\bibinfo {year} {2020})}\BibitemShut
  {NoStop}%
\bibitem [{\citenamefont {Ghiotto}\ \emph {et~al.}(2021)\citenamefont
  {Ghiotto}, \citenamefont {Shih}, \citenamefont {Pereira}, \citenamefont
  {Rhodes}, \citenamefont {Kim}, \citenamefont {Zang}, \citenamefont {Millis},
  \citenamefont {Watanabe}, \citenamefont {Taniguchi}, \citenamefont {Hone}
  \emph {et~al.}}]{ghiotto2021quantum}%
  \BibitemOpen
  \bibfield  {author} {\bibinfo {author} {\bibfnamefont {A.}~\bibnamefont
  {Ghiotto}}, \bibinfo {author} {\bibfnamefont {E.-M.}\ \bibnamefont {Shih}},
  \bibinfo {author} {\bibfnamefont {G.~S.}\ \bibnamefont {Pereira}}, \bibinfo
  {author} {\bibfnamefont {D.~A.}\ \bibnamefont {Rhodes}}, \bibinfo {author}
  {\bibfnamefont {B.}~\bibnamefont {Kim}}, \bibinfo {author} {\bibfnamefont
  {J.}~\bibnamefont {Zang}}, \bibinfo {author} {\bibfnamefont {A.~J.}\
  \bibnamefont {Millis}}, \bibinfo {author} {\bibfnamefont {K.}~\bibnamefont
  {Watanabe}}, \bibinfo {author} {\bibfnamefont {T.}~\bibnamefont {Taniguchi}},
  \bibinfo {author} {\bibfnamefont {J.~C.}\ \bibnamefont {Hone}}, \emph
  {et~al.},\ }\href@noop {} {\bibfield  {journal} {\bibinfo  {journal}
  {Nature}\ }\textbf {\bibinfo {volume} {597}},\ \bibinfo {pages} {345}
  (\bibinfo {year} {2021})}\BibitemShut {NoStop}%
\bibitem [{\citenamefont {Bi}\ and\ \citenamefont
  {Fu}(2021)}]{bi2021excitonic}%
  \BibitemOpen
  \bibfield  {author} {\bibinfo {author} {\bibfnamefont {Z.}~\bibnamefont
  {Bi}}\ and\ \bibinfo {author} {\bibfnamefont {L.}~\bibnamefont {Fu}},\
  }\href@noop {} {\bibfield  {journal} {\bibinfo  {journal} {Nature
  communications}\ }\textbf {\bibinfo {volume} {12}},\ \bibinfo {pages} {1}
  (\bibinfo {year} {2021})}\BibitemShut {NoStop}%
\bibitem [{\citenamefont {Zang}\ \emph {et~al.}(2021)\citenamefont {Zang},
  \citenamefont {Wang}, \citenamefont {Cano},\ and\ \citenamefont
  {Millis}}]{zang2021hartree}%
  \BibitemOpen
  \bibfield  {author} {\bibinfo {author} {\bibfnamefont {J.}~\bibnamefont
  {Zang}}, \bibinfo {author} {\bibfnamefont {J.}~\bibnamefont {Wang}}, \bibinfo
  {author} {\bibfnamefont {J.}~\bibnamefont {Cano}},\ and\ \bibinfo {author}
  {\bibfnamefont {A.~J.}\ \bibnamefont {Millis}},\ }\href@noop {} {\bibfield
  {journal} {\bibinfo  {journal} {Physical Review B}\ }\textbf {\bibinfo
  {volume} {104}},\ \bibinfo {pages} {075150} (\bibinfo {year}
  {2021})}\BibitemShut {NoStop}%
\bibitem [{\citenamefont {Zang}\ \emph {et~al.}(2022)\citenamefont {Zang},
  \citenamefont {Wang}, \citenamefont {Cano}, \citenamefont {Georges},\ and\
  \citenamefont {Millis}}]{zang2022dynamical}%
  \BibitemOpen
  \bibfield  {author} {\bibinfo {author} {\bibfnamefont {J.}~\bibnamefont
  {Zang}}, \bibinfo {author} {\bibfnamefont {J.}~\bibnamefont {Wang}}, \bibinfo
  {author} {\bibfnamefont {J.}~\bibnamefont {Cano}}, \bibinfo {author}
  {\bibfnamefont {A.}~\bibnamefont {Georges}},\ and\ \bibinfo {author}
  {\bibfnamefont {A.~J.}\ \bibnamefont {Millis}},\ }\href@noop {} {\bibfield
  {journal} {\bibinfo  {journal} {Physical Review X}\ }\textbf {\bibinfo
  {volume} {12}},\ \bibinfo {pages} {021064} (\bibinfo {year}
  {2022})}\BibitemShut {NoStop}%
\bibitem [{\citenamefont {Devakul}\ \emph {et~al.}(2021)\citenamefont
  {Devakul}, \citenamefont {Cr{\'e}pel}, \citenamefont {Zhang},\ and\
  \citenamefont {Fu}}]{devakul2021magic}%
  \BibitemOpen
  \bibfield  {author} {\bibinfo {author} {\bibfnamefont {T.}~\bibnamefont
  {Devakul}}, \bibinfo {author} {\bibfnamefont {V.}~\bibnamefont {Cr{\'e}pel}},
  \bibinfo {author} {\bibfnamefont {Y.}~\bibnamefont {Zhang}},\ and\ \bibinfo
  {author} {\bibfnamefont {L.}~\bibnamefont {Fu}},\ }\href@noop {} {\bibfield
  {journal} {\bibinfo  {journal} {Nature communications}\ }\textbf {\bibinfo
  {volume} {12}},\ \bibinfo {pages} {1} (\bibinfo {year} {2021})}\BibitemShut
  {NoStop}%
\bibitem [{\citenamefont {Li}\ \emph {et~al.}(2021{\natexlab{c}})\citenamefont
  {Li}, \citenamefont {Kumar}, \citenamefont {Sun},\ and\ \citenamefont
  {Lin}}]{li2021spontaneous}%
  \BibitemOpen
  \bibfield  {author} {\bibinfo {author} {\bibfnamefont {H.}~\bibnamefont
  {Li}}, \bibinfo {author} {\bibfnamefont {U.}~\bibnamefont {Kumar}}, \bibinfo
  {author} {\bibfnamefont {K.}~\bibnamefont {Sun}},\ and\ \bibinfo {author}
  {\bibfnamefont {S.-Z.}\ \bibnamefont {Lin}},\ }\href@noop {} {\bibfield
  {journal} {\bibinfo  {journal} {Physical Review Research}\ }\textbf {\bibinfo
  {volume} {3}},\ \bibinfo {pages} {L032070} (\bibinfo {year}
  {2021}{\natexlab{c}})}\BibitemShut {NoStop}%
\bibitem [{sup()}]{suppmat}%
  \BibitemOpen
  \href {https://publisherwebsite.com/doi} {}\bibinfo {note} {See supplementary
  materials at URL}\BibitemShut {NoStop}%
\bibitem [{\citenamefont {Qiu}\ \emph {et~al.}(2019)\citenamefont {Qiu},
  \citenamefont {Trushin}, \citenamefont {Fang}, \citenamefont {Verzhbitskiy},
  \citenamefont {Gao}, \citenamefont {Laksono}, \citenamefont {Yang},
  \citenamefont {Lyu}, \citenamefont {Li}, \citenamefont {Su} \emph
  {et~al.}}]{qiu2019giant}%
  \BibitemOpen
  \bibfield  {author} {\bibinfo {author} {\bibfnamefont {Z.}~\bibnamefont
  {Qiu}}, \bibinfo {author} {\bibfnamefont {M.}~\bibnamefont {Trushin}},
  \bibinfo {author} {\bibfnamefont {H.}~\bibnamefont {Fang}}, \bibinfo {author}
  {\bibfnamefont {I.}~\bibnamefont {Verzhbitskiy}}, \bibinfo {author}
  {\bibfnamefont {S.}~\bibnamefont {Gao}}, \bibinfo {author} {\bibfnamefont
  {E.}~\bibnamefont {Laksono}}, \bibinfo {author} {\bibfnamefont
  {M.}~\bibnamefont {Yang}}, \bibinfo {author} {\bibfnamefont {P.}~\bibnamefont
  {Lyu}}, \bibinfo {author} {\bibfnamefont {J.}~\bibnamefont {Li}}, \bibinfo
  {author} {\bibfnamefont {J.}~\bibnamefont {Su}}, \emph {et~al.},\ }\href@noop
  {} {\bibfield  {journal} {\bibinfo  {journal} {Science advances}\ }\textbf
  {\bibinfo {volume} {5}},\ \bibinfo {pages} {eaaw2347} (\bibinfo {year}
  {2019})}\BibitemShut {NoStop}%
\bibitem [{\citenamefont {Goodwin}\ \emph {et~al.}(2019)\citenamefont
  {Goodwin}, \citenamefont {Corsetti}, \citenamefont {Mostofi},\ and\
  \citenamefont {Lischner}}]{goodwin2019twist}%
  \BibitemOpen
  \bibfield  {author} {\bibinfo {author} {\bibfnamefont {Z.~A.}\ \bibnamefont
  {Goodwin}}, \bibinfo {author} {\bibfnamefont {F.}~\bibnamefont {Corsetti}},
  \bibinfo {author} {\bibfnamefont {A.~A.}\ \bibnamefont {Mostofi}},\ and\
  \bibinfo {author} {\bibfnamefont {J.}~\bibnamefont {Lischner}},\ }\href@noop
  {} {\bibfield  {journal} {\bibinfo  {journal} {Physical Review B}\ }\textbf
  {\bibinfo {volume} {100}},\ \bibinfo {pages} {121106} (\bibinfo {year}
  {2019})}\BibitemShut {NoStop}%
\bibitem [{\citenamefont {Kim}\ \emph {et~al.}(2020)\citenamefont {Kim},
  \citenamefont {Xu}, \citenamefont {Berdyugin}, \citenamefont {Principi},
  \citenamefont {Slizovskiy}, \citenamefont {Xin}, \citenamefont
  {Kumaravadivel}, \citenamefont {Kuang}, \citenamefont {Hamer}, \citenamefont
  {Krishna~Kumar} \emph {et~al.}}]{kim2020control}%
  \BibitemOpen
  \bibfield  {author} {\bibinfo {author} {\bibfnamefont {M.}~\bibnamefont
  {Kim}}, \bibinfo {author} {\bibfnamefont {S.}~\bibnamefont {Xu}}, \bibinfo
  {author} {\bibfnamefont {A.}~\bibnamefont {Berdyugin}}, \bibinfo {author}
  {\bibfnamefont {A.}~\bibnamefont {Principi}}, \bibinfo {author}
  {\bibfnamefont {S.}~\bibnamefont {Slizovskiy}}, \bibinfo {author}
  {\bibfnamefont {N.}~\bibnamefont {Xin}}, \bibinfo {author} {\bibfnamefont
  {P.}~\bibnamefont {Kumaravadivel}}, \bibinfo {author} {\bibfnamefont
  {W.}~\bibnamefont {Kuang}}, \bibinfo {author} {\bibfnamefont
  {M.}~\bibnamefont {Hamer}}, \bibinfo {author} {\bibfnamefont
  {R.}~\bibnamefont {Krishna~Kumar}}, \emph {et~al.},\ }\href@noop {}
  {\bibfield  {journal} {\bibinfo  {journal} {Nature communications}\ }\textbf
  {\bibinfo {volume} {11}},\ \bibinfo {pages} {1} (\bibinfo {year}
  {2020})}\BibitemShut {NoStop}%
\bibitem [{\citenamefont {Xu}\ \emph {et~al.}(2022)\citenamefont {Xu},
  \citenamefont {Kang}, \citenamefont {Watanabe}, \citenamefont {Taniguchi},
  \citenamefont {Mak},\ and\ \citenamefont {Shan}}]{xu2022tunable}%
  \BibitemOpen
  \bibfield  {author} {\bibinfo {author} {\bibfnamefont {Y.}~\bibnamefont
  {Xu}}, \bibinfo {author} {\bibfnamefont {K.}~\bibnamefont {Kang}}, \bibinfo
  {author} {\bibfnamefont {K.}~\bibnamefont {Watanabe}}, \bibinfo {author}
  {\bibfnamefont {T.}~\bibnamefont {Taniguchi}}, \bibinfo {author}
  {\bibfnamefont {K.~F.}\ \bibnamefont {Mak}},\ and\ \bibinfo {author}
  {\bibfnamefont {J.}~\bibnamefont {Shan}},\ }\href@noop {} {\bibfield
  {journal} {\bibinfo  {journal} {arXiv preprint arXiv:2202.02055}\ } (\bibinfo
  {year} {2022})}\BibitemShut {NoStop}%
\bibitem [{\citenamefont {Alavirad}\ and\ \citenamefont
  {Sau}(2020)}]{alavirad2020ferromagnetism}%
  \BibitemOpen
  \bibfield  {author} {\bibinfo {author} {\bibfnamefont {Y.}~\bibnamefont
  {Alavirad}}\ and\ \bibinfo {author} {\bibfnamefont {J.}~\bibnamefont {Sau}},\
  }\href@noop {} {\bibfield  {journal} {\bibinfo  {journal} {Physical Review
  B}\ }\textbf {\bibinfo {volume} {102}},\ \bibinfo {pages} {235123} (\bibinfo
  {year} {2020})}\BibitemShut {NoStop}%
\bibitem [{\citenamefont {Haldane}(1988)}]{haldane1988model}%
  \BibitemOpen
  \bibfield  {author} {\bibinfo {author} {\bibfnamefont {F.~D.~M.}\
  \bibnamefont {Haldane}},\ }\href@noop {} {\bibfield  {journal} {\bibinfo
  {journal} {Physical review letters}\ }\textbf {\bibinfo {volume} {61}},\
  \bibinfo {pages} {2015} (\bibinfo {year} {1988})}\BibitemShut {NoStop}%
\bibitem [{\citenamefont {Wu}\ \emph {et~al.}(2012)\citenamefont {Wu},
  \citenamefont {Bernevig},\ and\ \citenamefont {Regnault}}]{wu2012zoology}%
  \BibitemOpen
  \bibfield  {author} {\bibinfo {author} {\bibfnamefont {Y.-L.}\ \bibnamefont
  {Wu}}, \bibinfo {author} {\bibfnamefont {B.~A.}\ \bibnamefont {Bernevig}},\
  and\ \bibinfo {author} {\bibfnamefont {N.}~\bibnamefont {Regnault}},\
  }\href@noop {} {\bibfield  {journal} {\bibinfo  {journal} {Physical Review
  B}\ }\textbf {\bibinfo {volume} {85}},\ \bibinfo {pages} {075116} (\bibinfo
  {year} {2012})}\BibitemShut {NoStop}%
\bibitem [{\citenamefont {Dobard{\v{z}}i{\'c}}\ \emph
  {et~al.}(2013)\citenamefont {Dobard{\v{z}}i{\'c}}, \citenamefont
  {Milovanovi{\'c}},\ and\ \citenamefont
  {Regnault}}]{dobardvzic2013geometrical}%
  \BibitemOpen
  \bibfield  {author} {\bibinfo {author} {\bibfnamefont {E.}~\bibnamefont
  {Dobard{\v{z}}i{\'c}}}, \bibinfo {author} {\bibfnamefont {M.}~\bibnamefont
  {Milovanovi{\'c}}},\ and\ \bibinfo {author} {\bibfnamefont {N.}~\bibnamefont
  {Regnault}},\ }\href@noop {} {\bibfield  {journal} {\bibinfo  {journal}
  {Physical Review B}\ }\textbf {\bibinfo {volume} {88}},\ \bibinfo {pages}
  {115117} (\bibinfo {year} {2013})}\BibitemShut {NoStop}%
\bibitem [{\citenamefont {Bernevig}\ and\ \citenamefont
  {Regnault}(2012)}]{bernevig2012emergent}%
  \BibitemOpen
  \bibfield  {author} {\bibinfo {author} {\bibfnamefont {B.~A.}\ \bibnamefont
  {Bernevig}}\ and\ \bibinfo {author} {\bibfnamefont {N.}~\bibnamefont
  {Regnault}},\ }\href@noop {} {\bibfield  {journal} {\bibinfo  {journal}
  {Physical Review B}\ }\textbf {\bibinfo {volume} {85}},\ \bibinfo {pages}
  {075128} (\bibinfo {year} {2012})}\BibitemShut {NoStop}%
\bibitem [{\citenamefont {Grushin}\ \emph {et~al.}(2015)\citenamefont
  {Grushin}, \citenamefont {Motruk}, \citenamefont {Zaletel},\ and\
  \citenamefont {Pollmann}}]{grushin2015characterization}%
  \BibitemOpen
  \bibfield  {author} {\bibinfo {author} {\bibfnamefont {A.~G.}\ \bibnamefont
  {Grushin}}, \bibinfo {author} {\bibfnamefont {J.}~\bibnamefont {Motruk}},
  \bibinfo {author} {\bibfnamefont {M.~P.}\ \bibnamefont {Zaletel}},\ and\
  \bibinfo {author} {\bibfnamefont {F.}~\bibnamefont {Pollmann}},\ }\href@noop
  {} {\bibfield  {journal} {\bibinfo  {journal} {Physical Review B}\ }\textbf
  {\bibinfo {volume} {91}},\ \bibinfo {pages} {035136} (\bibinfo {year}
  {2015})}\BibitemShut {NoStop}%
\bibitem [{\citenamefont {Regnault}\ and\ \citenamefont
  {Bernevig}(2011)}]{regnault2011fractional}%
  \BibitemOpen
  \bibfield  {author} {\bibinfo {author} {\bibfnamefont {N.}~\bibnamefont
  {Regnault}}\ and\ \bibinfo {author} {\bibfnamefont {B.~A.}\ \bibnamefont
  {Bernevig}},\ }\href@noop {} {\bibfield  {journal} {\bibinfo  {journal}
  {Physical Review X}\ }\textbf {\bibinfo {volume} {1}},\ \bibinfo {pages}
  {021014} (\bibinfo {year} {2011})}\BibitemShut {NoStop}%
\bibitem [{\citenamefont {Sheng}\ \emph {et~al.}(2011)\citenamefont {Sheng},
  \citenamefont {Gu}, \citenamefont {Sun},\ and\ \citenamefont
  {Sheng}}]{sheng2011fractional}%
  \BibitemOpen
  \bibfield  {author} {\bibinfo {author} {\bibfnamefont {D.}~\bibnamefont
  {Sheng}}, \bibinfo {author} {\bibfnamefont {Z.-C.}\ \bibnamefont {Gu}},
  \bibinfo {author} {\bibfnamefont {K.}~\bibnamefont {Sun}},\ and\ \bibinfo
  {author} {\bibfnamefont {L.}~\bibnamefont {Sheng}},\ }\href@noop {}
  {\bibfield  {journal} {\bibinfo  {journal} {Nature communications}\ }\textbf
  {\bibinfo {volume} {2}},\ \bibinfo {pages} {1} (\bibinfo {year}
  {2011})}\BibitemShut {NoStop}%
\bibitem [{\citenamefont {Neupert}\ \emph {et~al.}(2011)\citenamefont
  {Neupert}, \citenamefont {Santos}, \citenamefont {Chamon},\ and\
  \citenamefont {Mudry}}]{neupert2011fractional}%
  \BibitemOpen
  \bibfield  {author} {\bibinfo {author} {\bibfnamefont {T.}~\bibnamefont
  {Neupert}}, \bibinfo {author} {\bibfnamefont {L.}~\bibnamefont {Santos}},
  \bibinfo {author} {\bibfnamefont {C.}~\bibnamefont {Chamon}},\ and\ \bibinfo
  {author} {\bibfnamefont {C.}~\bibnamefont {Mudry}},\ }\href@noop {}
  {\bibfield  {journal} {\bibinfo  {journal} {Physical review letters}\
  }\textbf {\bibinfo {volume} {106}},\ \bibinfo {pages} {236804} (\bibinfo
  {year} {2011})}\BibitemShut {NoStop}%
\bibitem [{\citenamefont {Cr{\'e}pel}\ \emph {et~al.}(2020)\citenamefont
  {Cr{\'e}pel}, \citenamefont {Estienne},\ and\ \citenamefont
  {Regnault}}]{crepel2020microscopic}%
  \BibitemOpen
  \bibfield  {author} {\bibinfo {author} {\bibfnamefont {V.}~\bibnamefont
  {Cr{\'e}pel}}, \bibinfo {author} {\bibfnamefont {B.}~\bibnamefont
  {Estienne}},\ and\ \bibinfo {author} {\bibfnamefont {N.}~\bibnamefont
  {Regnault}},\ }\href@noop {} {\bibfield  {journal} {\bibinfo  {journal}
  {Physical Review B}\ }\textbf {\bibinfo {volume} {101}},\ \bibinfo {pages}
  {235158} (\bibinfo {year} {2020})}\BibitemShut {NoStop}%
\bibitem [{\citenamefont {Cr{\'e}pel}\ \emph {et~al.}(2018)\citenamefont
  {Cr{\'e}pel}, \citenamefont {Estienne}, \citenamefont {Bernevig},
  \citenamefont {Lecheminant},\ and\ \citenamefont
  {Regnault}}]{crepel2018matrix}%
  \BibitemOpen
  \bibfield  {author} {\bibinfo {author} {\bibfnamefont {V.}~\bibnamefont
  {Cr{\'e}pel}}, \bibinfo {author} {\bibfnamefont {B.}~\bibnamefont
  {Estienne}}, \bibinfo {author} {\bibfnamefont {B.~A.}\ \bibnamefont
  {Bernevig}}, \bibinfo {author} {\bibfnamefont {P.}~\bibnamefont
  {Lecheminant}},\ and\ \bibinfo {author} {\bibfnamefont {N.}~\bibnamefont
  {Regnault}},\ }\href@noop {} {\bibfield  {journal} {\bibinfo  {journal}
  {Physical Review B}\ }\textbf {\bibinfo {volume} {97}},\ \bibinfo {pages}
  {165136} (\bibinfo {year} {2018})}\BibitemShut {NoStop}%
\bibitem [{\citenamefont {Grushin}\ \emph {et~al.}(2012)\citenamefont
  {Grushin}, \citenamefont {Neupert}, \citenamefont {Chamon},\ and\
  \citenamefont {Mudry}}]{grushin2012enhancing}%
  \BibitemOpen
  \bibfield  {author} {\bibinfo {author} {\bibfnamefont {A.~G.}\ \bibnamefont
  {Grushin}}, \bibinfo {author} {\bibfnamefont {T.}~\bibnamefont {Neupert}},
  \bibinfo {author} {\bibfnamefont {C.}~\bibnamefont {Chamon}},\ and\ \bibinfo
  {author} {\bibfnamefont {C.}~\bibnamefont {Mudry}},\ }\href@noop {}
  {\bibfield  {journal} {\bibinfo  {journal} {Physical Review B}\ }\textbf
  {\bibinfo {volume} {86}},\ \bibinfo {pages} {205125} (\bibinfo {year}
  {2012})}\BibitemShut {NoStop}%
\bibitem [{\citenamefont {Neupert}\ \emph {et~al.}(2012)\citenamefont
  {Neupert}, \citenamefont {Santos}, \citenamefont {Chamon},\ and\
  \citenamefont {Mudry}}]{neupert2012elementary}%
  \BibitemOpen
  \bibfield  {author} {\bibinfo {author} {\bibfnamefont {T.}~\bibnamefont
  {Neupert}}, \bibinfo {author} {\bibfnamefont {L.}~\bibnamefont {Santos}},
  \bibinfo {author} {\bibfnamefont {C.}~\bibnamefont {Chamon}},\ and\ \bibinfo
  {author} {\bibfnamefont {C.}~\bibnamefont {Mudry}},\ }\href@noop {}
  {\bibfield  {journal} {\bibinfo  {journal} {Physical Review B}\ }\textbf
  {\bibinfo {volume} {86}},\ \bibinfo {pages} {165133} (\bibinfo {year}
  {2012})}\BibitemShut {NoStop}%
\bibitem [{Note1()}]{Note1}%
  \BibitemOpen
  \bibinfo {note} {In this calculation, both the Berry curvature $B_{\protect
  \bm {k}}$ and the Chern number $C_\uparrow $ were computed with the finite
  momentum discretization imposed by the $6 \times N_2$ lattice, using the
  method of Ref.~\cite {fukui2005chern}.}\BibitemShut {Stop}%
\bibitem [{\citenamefont {Cr{\'e}pel}\ \emph {et~al.}(2019)\citenamefont
  {Cr{\'e}pel}, \citenamefont {Regnault},\ and\ \citenamefont
  {Estienne}}]{crepel2019matrix}%
  \BibitemOpen
  \bibfield  {author} {\bibinfo {author} {\bibfnamefont {V.}~\bibnamefont
  {Cr{\'e}pel}}, \bibinfo {author} {\bibfnamefont {N.}~\bibnamefont
  {Regnault}},\ and\ \bibinfo {author} {\bibfnamefont {B.}~\bibnamefont
  {Estienne}},\ }\href@noop {} {\bibfield  {journal} {\bibinfo  {journal}
  {Physical Review B}\ }\textbf {\bibinfo {volume} {100}},\ \bibinfo {pages}
  {125128} (\bibinfo {year} {2019})}\BibitemShut {NoStop}%
\bibitem [{Note2()}]{Note2}%
  \BibitemOpen
  \bibinfo {note} {While the gap $\Delta $ appears robust in our calculations,
  a finite size scaling of the gap to the thermodynamics is however plagued by
  commensuration effects stabilizing CDW when both $N_1$ and $N_2$ are
  multiples of six (Fig.~\ref {fig:FirstHintFCI}b). See Refs.~\cite
  {wu2012zoology} and~\cite {grushin2015characterization} for a longer
  discussion.}\BibitemShut {Stop}%
\bibitem [{\citenamefont {Thouless}(1989)}]{thouless1989level}%
  \BibitemOpen
  \bibfield  {author} {\bibinfo {author} {\bibfnamefont {D.}~\bibnamefont
  {Thouless}},\ }\href@noop {} {\bibfield  {journal} {\bibinfo  {journal}
  {Physical Review B}\ }\textbf {\bibinfo {volume} {40}},\ \bibinfo {pages}
  {12034} (\bibinfo {year} {1989})}\BibitemShut {NoStop}%
\bibitem [{\citenamefont {Zhang}\ \emph {et~al.}(2021)\citenamefont {Zhang},
  \citenamefont {Liu},\ and\ \citenamefont {Fu}}]{zhang2021electronic}%
  \BibitemOpen
  \bibfield  {author} {\bibinfo {author} {\bibfnamefont {Y.}~\bibnamefont
  {Zhang}}, \bibinfo {author} {\bibfnamefont {T.}~\bibnamefont {Liu}},\ and\
  \bibinfo {author} {\bibfnamefont {L.}~\bibnamefont {Fu}},\ }\href@noop {}
  {\bibfield  {journal} {\bibinfo  {journal} {Physical Review B}\ }\textbf
  {\bibinfo {volume} {103}},\ \bibinfo {pages} {155142} (\bibinfo {year}
  {2021})}\BibitemShut {NoStop}%
\bibitem [{\citenamefont {Kourtis}\ \emph {et~al.}(2014)\citenamefont
  {Kourtis}, \citenamefont {Neupert}, \citenamefont {Chamon},\ and\
  \citenamefont {Mudry}}]{kourtis2014fractional}%
  \BibitemOpen
  \bibfield  {author} {\bibinfo {author} {\bibfnamefont {S.}~\bibnamefont
  {Kourtis}}, \bibinfo {author} {\bibfnamefont {T.}~\bibnamefont {Neupert}},
  \bibinfo {author} {\bibfnamefont {C.}~\bibnamefont {Chamon}},\ and\ \bibinfo
  {author} {\bibfnamefont {C.}~\bibnamefont {Mudry}},\ }\href@noop {}
  {\bibfield  {journal} {\bibinfo  {journal} {Physical review letters}\
  }\textbf {\bibinfo {volume} {112}},\ \bibinfo {pages} {126806} (\bibinfo
  {year} {2014})}\BibitemShut {NoStop}%
\bibitem [{Note3()}]{Note3}%
  \BibitemOpen
  \bibinfo {note} {The inversion symmetry of the model makes states at momenta
  $(K_1,K_2)$ and $(-K_1,-K_2)$ coincide.}\BibitemShut {Stop}%
\bibitem [{\citenamefont {Xie}\ \emph {et~al.}(2021)\citenamefont {Xie},
  \citenamefont {Pierce}, \citenamefont {Park}, \citenamefont {Parker},
  \citenamefont {Khalaf}, \citenamefont {Ledwith}, \citenamefont {Cao},
  \citenamefont {Lee}, \citenamefont {Chen}, \citenamefont {Forrester} \emph
  {et~al.}}]{xie2021fractional}%
  \BibitemOpen
  \bibfield  {author} {\bibinfo {author} {\bibfnamefont {Y.}~\bibnamefont
  {Xie}}, \bibinfo {author} {\bibfnamefont {A.~T.}\ \bibnamefont {Pierce}},
  \bibinfo {author} {\bibfnamefont {J.~M.}\ \bibnamefont {Park}}, \bibinfo
  {author} {\bibfnamefont {D.~E.}\ \bibnamefont {Parker}}, \bibinfo {author}
  {\bibfnamefont {E.}~\bibnamefont {Khalaf}}, \bibinfo {author} {\bibfnamefont
  {P.}~\bibnamefont {Ledwith}}, \bibinfo {author} {\bibfnamefont
  {Y.}~\bibnamefont {Cao}}, \bibinfo {author} {\bibfnamefont {S.~H.}\
  \bibnamefont {Lee}}, \bibinfo {author} {\bibfnamefont {S.}~\bibnamefont
  {Chen}}, \bibinfo {author} {\bibfnamefont {P.~R.}\ \bibnamefont {Forrester}},
  \emph {et~al.},\ }\href@noop {} {\bibfield  {journal} {\bibinfo  {journal}
  {Nature}\ }\textbf {\bibinfo {volume} {600}},\ \bibinfo {pages} {439}
  (\bibinfo {year} {2021})}\BibitemShut {NoStop}%
\bibitem [{\citenamefont {Xie}\ and\ \citenamefont
  {MacDonald}(2021)}]{xie2021weak}%
  \BibitemOpen
  \bibfield  {author} {\bibinfo {author} {\bibfnamefont {M.}~\bibnamefont
  {Xie}}\ and\ \bibinfo {author} {\bibfnamefont {A.~H.}\ \bibnamefont
  {MacDonald}},\ }\href@noop {} {\bibfield  {journal} {\bibinfo  {journal}
  {Physical Review Letters}\ }\textbf {\bibinfo {volume} {127}},\ \bibinfo
  {pages} {196401} (\bibinfo {year} {2021})}\BibitemShut {NoStop}%
\bibitem [{\citenamefont {Ledwith}\ \emph {et~al.}(2020)\citenamefont
  {Ledwith}, \citenamefont {Tarnopolsky}, \citenamefont {Khalaf},\ and\
  \citenamefont {Vishwanath}}]{ledwith2020fractional}%
  \BibitemOpen
  \bibfield  {author} {\bibinfo {author} {\bibfnamefont {P.~J.}\ \bibnamefont
  {Ledwith}}, \bibinfo {author} {\bibfnamefont {G.}~\bibnamefont
  {Tarnopolsky}}, \bibinfo {author} {\bibfnamefont {E.}~\bibnamefont
  {Khalaf}},\ and\ \bibinfo {author} {\bibfnamefont {A.}~\bibnamefont
  {Vishwanath}},\ }\href@noop {} {\bibfield  {journal} {\bibinfo  {journal}
  {Physical Review Research}\ }\textbf {\bibinfo {volume} {2}},\ \bibinfo
  {pages} {023237} (\bibinfo {year} {2020})}\BibitemShut {NoStop}%
\bibitem [{\citenamefont {Repellin}\ and\ \citenamefont
  {Senthil}(2020)}]{repellin2020chern}%
  \BibitemOpen
  \bibfield  {author} {\bibinfo {author} {\bibfnamefont {C.}~\bibnamefont
  {Repellin}}\ and\ \bibinfo {author} {\bibfnamefont {T.}~\bibnamefont
  {Senthil}},\ }\href@noop {} {\bibfield  {journal} {\bibinfo  {journal}
  {Physical Review Research}\ }\textbf {\bibinfo {volume} {2}},\ \bibinfo
  {pages} {023238} (\bibinfo {year} {2020})}\BibitemShut {NoStop}%
\bibitem [{\citenamefont {Liu}\ \emph {et~al.}(2021)\citenamefont {Liu},
  \citenamefont {Abouelkomsan},\ and\ \citenamefont {Bergholtz}}]{liu2021gate}%
  \BibitemOpen
  \bibfield  {author} {\bibinfo {author} {\bibfnamefont {Z.}~\bibnamefont
  {Liu}}, \bibinfo {author} {\bibfnamefont {A.}~\bibnamefont {Abouelkomsan}},\
  and\ \bibinfo {author} {\bibfnamefont {E.~J.}\ \bibnamefont {Bergholtz}},\
  }\href@noop {} {\bibfield  {journal} {\bibinfo  {journal} {Physical Review
  Letters}\ }\textbf {\bibinfo {volume} {126}},\ \bibinfo {pages} {026801}
  (\bibinfo {year} {2021})}\BibitemShut {NoStop}%
\bibitem [{\citenamefont {Fukui}\ \emph {et~al.}(2005)\citenamefont {Fukui},
  \citenamefont {Hatsugai},\ and\ \citenamefont {Suzuki}}]{fukui2005chern}%
  \BibitemOpen
  \bibfield  {author} {\bibinfo {author} {\bibfnamefont {T.}~\bibnamefont
  {Fukui}}, \bibinfo {author} {\bibfnamefont {Y.}~\bibnamefont {Hatsugai}},\
  and\ \bibinfo {author} {\bibfnamefont {H.}~\bibnamefont {Suzuki}},\
  }\href@noop {} {\bibfield  {journal} {\bibinfo  {journal} {Journal of the
  Physical Society of Japan}\ }\textbf {\bibinfo {volume} {74}},\ \bibinfo
  {pages} {1674} (\bibinfo {year} {2005})}\BibitemShut {NoStop}%
\end{thebibliography}%

\onecolumngrid
\newpage
\makeatletter 

\begin{center}
\textbf{\large Supplementary material for `` \@title ''} \\[10pt]
Valentin Cr\'epel and Liang Fu \\
\textit{Department of Physics, Massachusetts Institute of Technology, Cambridge, Massachusetts 02139, USA}
\end{center}
\vspace{20pt}

\setcounter{figure}{0}
\setcounter{section}{0}
\setcounter{equation}{0}

\renewcommand{\thefigure}{S\@arabic\c@figure}
\renewcommand{\thesection}{S-\@Roman\c@section}
\makeatother

\twocolumngrid

\appendix

\newpage

\section{Continuum model} \label{app:continuummodel}

\begin{figure*}
\centering
\includegraphics[width=\textwidth]{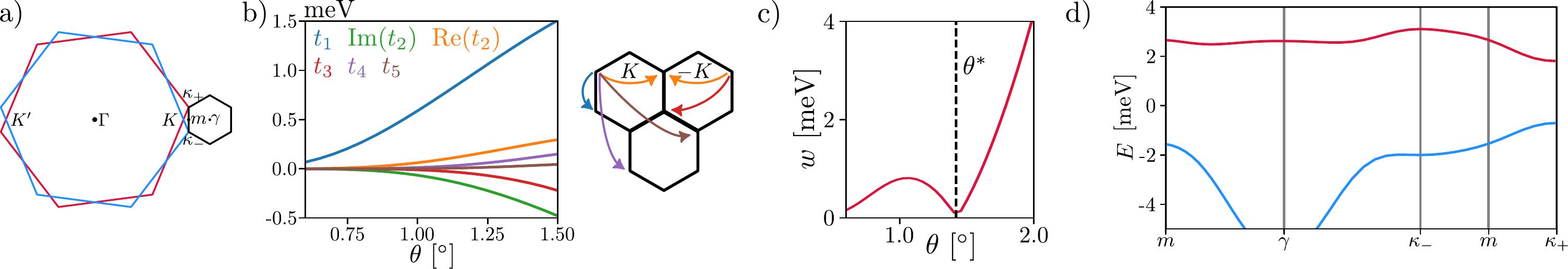}
\caption{a) Brillouin zone of the original top and bottom layer (blue and red), and moiré Brillouin zone (black). b) Tight-binding parameters reproduced from Ref.~\cite{devakul2021magic}, the $n$-th nearest neighbors are shown on the right. The second nearest neighbor tunneling $t_2$ is equal in magnitude, but have opposite phases for electrons of the $\pm K$ valleys. c) The bandwidth of the topmost moiré valence band against twist angle, showing a magic-angle $\theta^*=1.45^\circ$. d) Band structure obtained from the continuum and a layer potential difference $M=1.5$meV.  
}
\label{figapp:tightbinding}
\end{figure*}

As detailed in the main text, we study twisted TMD homobilayers, where the top layer is rotated by $\theta$ with respect the bottom layer around a point where the metallic atoms of the two layers coincide. Due to the spin/valley locking, each monolayer exhibits two valleys, one at $+K$ with spin-$\uparrow$ and its time-reversal conjugate at $-K$ with spin-$\downarrow$ (see Fig.~\ref{figapp:tightbinding}a). The angle $\theta$ between the two layers slightly displaces the $K$ points of the original monolayers, which become the two corners $\kappa_\pm$ of the moiré Brillouin zone (mBZ).

This moiré structure, reinforced by large lattice relaxation effects~\cite{devakul2021magic}, induces slowly varying scalar potentials $V_{\pm}(\vec{r})$ in each layers and interlayer tunneling $T(\vec{r})$. These potentials hybridize the $\pm K$ valleys of the two layers and yield a set of moiré bands, which can be captured by the minimal continuum model established in Refs.~\cite{wu2019topological,devakul2021magic}:
\begin{equation} \label{appeq:ContinuumModel}
\mathcal{H}_\uparrow (\vec{k}) = \begin{pmatrix} & - \frac{|\vec{k}-\vec{\kappa}_+|^2}{2m^*} + V_+(\vec{r}) & T(\vec{r}) \\ &
T^\dagger (\vec{r}) & - \frac{|\vec{k}-\vec{\kappa}_-|^2}{2m^*} + V_-(\vec{r})
\end{pmatrix} ,
\end{equation}
for spin/valley $\uparrow$ holes, where $m^*$ is the effective hole mass in the original monolayers. 
$\mathcal{H}_\downarrow$ is obtained by time-reversal conjugation. 
Here, the scalar potentials $V_\pm$ and inter-layer tunneling $T$ are approximated by their lowest harmonics on the moiré lattice
\begin{subequations} \begin{eqnarray}
V_\pm (\vec{r}) &=& 2V \sum_{j=1,3,5} \cos( \vec{g}_j \cdot \vec{r} \pm \psi) , \\
T(\vec{r}) &=& w \left( 1 + e^{-i \vec{g}_2 \cdot \vec{r}} + e^{-i \vec{g}_3 \cdot \vec{r}} \right) ,
\end{eqnarray} \end{subequations}
with $\vec{g}_j$ the counter-clockwise rotations of the moiré reciprocal lattice vector $\vec{g}_1 = (4\pi\theta / \sqrt{3} a_0, 0)$, and $a_0$ the monolayer lattice constant. 
The parameters $(V,w,\psi)$ are determined from large-scale DFT calculations. 
In this article, we focus on WSe$_2$ homobilayers, for which $V = \SI{9}{\milli\electronvolt}$, $w = \SI{18}{\milli\electronvolt}$ and $\psi = 128^\circ$~\cite{devakul2021magic}.

For twist angles $\theta < \theta_{\rm KM} = 1.51^\circ$, the topmost two bands of Eq.~\ref{appeq:ContinuumModel} have opposite Chern number, $+1$ and $-1$, depicted in Fig.~\ref{fig:ContinuumModel}b for $\theta = 1.2^\circ$. As a consequence, these bands can be described by a tight-binding model with localized Wannier orbitals. These orbitals are shown to be centered at the MX and XM regions, thus forming a honeycomb lattice (see inset of Fig.~\ref{fig:ContinuumModel}). The effective tight-binding model was derived by projecting $\mathcal{H}_\uparrow$ on this Wannier basis. It corresponds to a generalized Kane-Mele model with long range tunnelings Eq.~\ref{eq:LongrangeKaneMele}. The parameters $t_n$ are reproduced from Ref.~\cite{devakul2021magic} in Fig.~\ref{figapp:tightbinding}b, with the convention that $\langle \vec{r}, \vec{r}' \rangle_2$ denotes a link where the path $\vec{r} \to \vec{r}'$ turns right.
We observe that $t_5$ is negligible for all twist angle below $\theta_{\rm KM}$, which justify the above truncation. Plotting the bandwidth of the topmost valence band as a function of the twist angle reveals the magic-angle introduced in the main text, see Fig.~\ref{figapp:tightbinding}c.

Finally, we can also account for the application of an out-of-plane magnetic field (as the one used in the main text) by adding a constant but opposite term $\pm M/2$ to each of the layer. As long as $M<\SI{1.5}{\milli\electronvolt}$, the gap between the two first band remains larger than the bandwidth of the topmost valence band, as illustrated in Fig.~\ref{figapp:tightbinding}d.

\section{Spin-flip spectrum} \label{app:Magnon}

In the limit of perfectly flat band at unit filling and large $U$, the ground state of the system is a ferromagnetic band insulator with the topmost valence band fully filled with holes of a single spin/valley specie. In this appendix, we determine the boundary of this ferromagnetic phase as a function of total density $n \leq 1$ and twist angle $\theta$ using a spin-flip analysis. More precisely, we compute the magnon spectrum above the spin-polarized ground state and locate the boundary of the ferromagnetic phase as the points where the minimum of the magnon spectrum reaches zero energy.

The most relevant energy scale in the problem is the on-site interaction $U$. Keeping only this term, the ferromagnetic state $\ket{\Psi} = \prod_{\vec{k}}'  c_{\vec{k},-,\uparrow}^\dagger \ket{\emptyset}$, obtained by filling the $N$ lowest orbitals with spin/valley $\uparrow$, is an exact eigenstate of the interacting problem. A necessary condition for this Slater-type ferromagnetic state to be the ground state of the full problem is that spin-flip created above it should have a gapped spectrum. To check this, we diagonalize the full interacting Hamiltonian in the basis 
\begin{equation}
\ket{\vec{Q}, \vec{k}, n=\pm} = c_{\vec{Q} + \vec{k}, n, \downarrow}^\dagger  c_{\vec{k}, -, \uparrow} \ket{\Psi} , 
\end{equation}
where $\vec{Q}$ is a good quantum number, while $\vec{k}$ runs over the occupied orbitals and $n=\pm$ spans the two bands of the model. The matrix elements between these Slater-determinant states read
\begin{widetext}
\begin{align} \label{appeq_SpinWaveAnalysis} 
& \braOket{\vec{Q}, \vec{k}', n'}{\mathcal{H}+\mathcal{H}_{\rm int}}{\vec{Q}, \vec{k}, n} = \delta_{n,n'} \delta_{\vec{k},\vec{k}'} E_{\rm ferro}  + \delta_{n,n'}\delta_{\vec{k},\vec{k}'} \left( \varepsilon_{\vec{Q}+\vec{k},n,\downarrow}
- \varepsilon_{\vec{k},n,\uparrow}
\right)  \\ 
& + \delta_{\vec{k},\vec{k}'} \left[ U n_{\rm ferro}^A u_{\vec{k}+\vec{Q},n',\downarrow}^{A \; *} u_{\vec{k}+\vec{Q},n,\downarrow}^A + U n_{\rm ferro}^B  u_{\vec{k}+\vec{Q},n',\downarrow}^{B \; *} u_{\vec{k}+\vec{Q},n,\downarrow}^B \right]
- \frac{U}{N_1N_2} \sum_{\tau = A/B} \left( u_{\vec{k}',-,\uparrow}^{\tau \; *} u_{\vec{k}'+\vec{Q},n',\downarrow}^\tau \right)^* \left( u_{\vec{k},-,\uparrow}^{\tau \; *} u_{\vec{k}+\vec{Q},n,\downarrow}^\tau \right) , \notag 
\end{align}
\end{widetext}
with $u_{\vec{k},n,\sigma}^\tau$ the weight of the Bloch orbitals of the $n$-th band with momentum $\vec{k}$ spin $\sigma$ -- with energy $\varepsilon_{\vec{k},n,\sigma}$ -- on the sublattice $\tau$. $E_{\rm ferro}$ is the total energy of $\ket{\Psi}$, and $n_{\rm ferro}^\tau$ denotes its total density on sublattice $\tau = A/B$. The three lines in Eq.~\ref{appeq_SpinWaveAnalysis} can be respectively interpreted as the single-particle, the Hartree and the Fock contributions to the magnon energy. We tested this spin-flip analysis by reproducing the results of Ref.~\cite{devakul2021magic} (Fig.~5).

\begin{figure}
\centering
\includegraphics[width=\columnwidth]{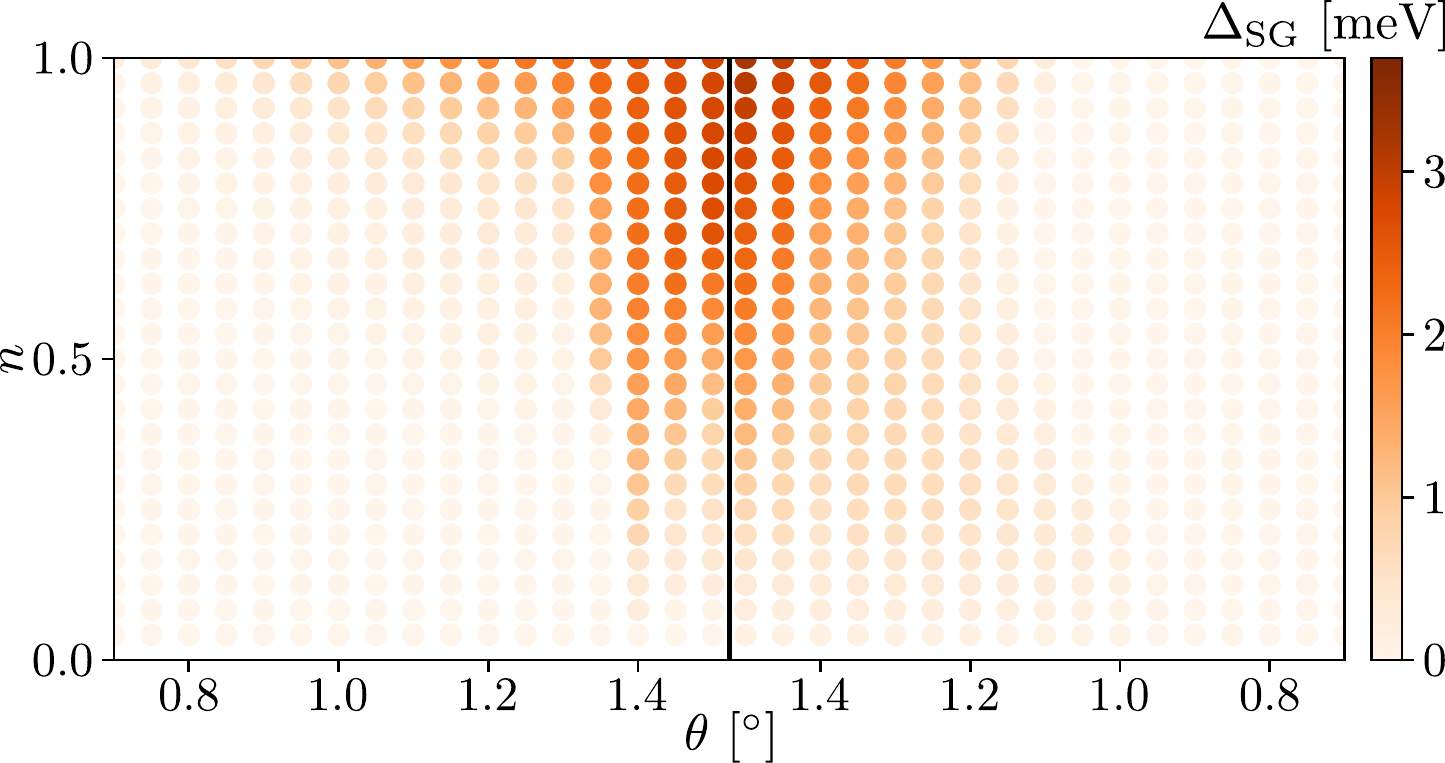}
\caption{The spin gap obtained from the spin-flip analysis detailed in App.~\ref{app:Magnon} for the two-band generalized tight-binding model Eq.~\ref{eq:LongrangeKaneMele} (left) and for the continuum model Eq.~\ref{appeq:ContinuumModel} (right) qualitatively match.
}
\label{figapp:ComparisonContinuumKM}
\end{figure}

In a similar manner, we can study the stability of the ferromegnetic state at the level of the continuum model directly. Since our analysis mostly focus on the small twist-angle limit where the two band model of App.~\ref{app:continuummodel} faithfully capture the band structure, a complete filling-dependent Hartree-Fock account of Coulomb interaction at twist angle $\theta >1.5^\circ$ goes beyond the scope of our analysis. We can nevertheless consider the effect of the most relevant part of the Coulomb interaction with contact terms, as we did above for the generalized Kane-Mele. The only differences with Eq.~\ref{appeq_SpinWaveAnalysis} is that the sublattice indices $A/B$ should be replaced by the top/bottom layer index, and all products of of the form $u_{\vec{k}}^* u_{\vec{k}'} \to \sum_{\vec{G}, \vec{G}'} u_{\vec{k}+\vec{G}}^* u_{\vec{k}'+\vec{G}'}$ to account for the Brillouin zone folding and the emergence of the moiré structure. The on-site interaction strength $U = \SI{25}{\milli\electronvolt}$ used in the tight-binding model is a combination of the Wannier orbital widths and the short-range Coulomb repulsion $U_{\rm cont}$ of both layers. We set the value of $U_{\rm cont} \simeq \SI{12.5}{\milli\electronvolt}$ to qualitatively match the results of our two band model in the range $0.6^\circ < \theta < 1.5^\circ$, as shown in Fig.~\ref{figapp:ComparisonContinuumKM}, and use this value of the interaction strength to study the ferromagnetic properties of the system at $\theta > 1.5^\circ$ in Fig.~\ref{fig:FMMphase} of the main text.

While restricting all interactions to an on-site Hubbard repulsion is well justified in our effective two-band model, especially for the very small twist angles and large moiré patterns under scrutiny, approximating the full Coulomb potential as an intra-layer local repulsion certainly is not for the continuum model. Therefore, our results for $\theta > 1.5^\circ$ should be understood as a qualitative extension of our analysis, showing that ferromegnatic behaviors remain so long as the topmost valence band is sufficiently flat and carries a non-zero spin/valley Chern number (see main text). A more complete account of interaction and polarization effects in twisted TMD will be the subject of a subsequent publication.

\section{1+2/3 filling} \label{app:OnThirdElectron}

\begin{figure}
\centering
\includegraphics[width=\columnwidth]{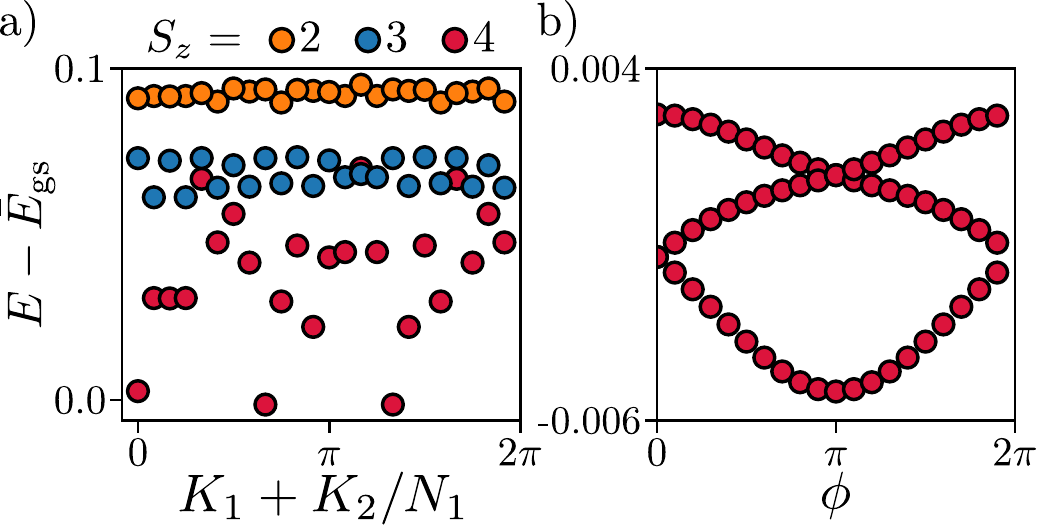}
\caption{a) Momentum resolved many-body spectrum obtained with band projected ED on the $6\times4$ lattice for $U=\SI{15}{\milli\electronvolt}$, $V = \SI{2}{\milli\electronvolt}$ and $\theta=1.4^\circ$. We only show the lowest energy states in each spin/momentum sector with energy less than $\SI{0.1}{\milli\electronvolt}$ above the ground state. 
b) The three nearly-degenerate ground states exhibit the characteristic spectral flow of FCI. 
All energies are given in meV.
}
\label{figapp:HighFilling}
\end{figure}

At $n=1+2/3$ filling, or equivalently 1/3 electron filling, of the topmost valence band, the ground state of the band projected Hamiltonian obtained with ED exhibits all characteristic features of the FQAH phase. As an example, we show in Fig.~\ref{figapp:HighFilling}a the many-body spectrum obtained at twist angle $\theta = 1.4^\circ$ with interaction parameters $U=\SI{15}{\milli\electronvolt}$ and  $V=\SI{2}{\milli\electronvolt}$. It has three nearly-degenerate ground states in the fully polarized sector, which displays the spectral flow of FCI (Fig.~\ref{figapp:HighFilling}b). While these evidence hint at a FQAH at larger carrier density $n=1+2/3$, we now argue that -- contrary to the $n=1/3$ case -- interband transitions are likely to weaken or destabilize this phase, especially for large on-site interaction strengths $U$.  

Indeed, the fully spin-polarized ground state at this filling has an on-site interaction energy $2U N/3$ due to the necessary double occupation of single particle states that occur at $n>1$. This energy can be released by promoting an electron from the spin-minority to an upper band and aligning it with the spin-majority. Such process costs an energy $\sim (\Delta - U/3)$, leading to a breakdown of ferromagnetism when $U$ is more than three times greater than the band gap. Thus, while $n=1+2/3$ is promising for detection of the FQAH in transport experiments, experiments might require substantial screening of the interaction to suppress interband spin-flip excitations.

\section{Transitions from FQAH to spin-polarized phases}  \label{app:FQAHneighbors}

We now study the phases directly adjacent to the FQAH observed in Fig.~\ref{fig:FirstHintFCI}, providing some guidance on what experimental knob to tune to realize the FQAH phase. Because the emergence of the FCI physics crucially relies on the small interaction energy of the three nearly-degenerate ground states, we expect that strong dielectric screening, \textit{e.g.} due to proximitized metallic gates, will be beneficial to the observation of FQAH physics in TMD bilayers. On the contrary, if the long range part of the Coulomb interaction is strong, the three degenerate ground states acquire some interaction energy comparable to those of states above the gap, which can lead to the destruction of the FQAH phase. 

\begin{figure}
\centering
\includegraphics[width=\columnwidth]{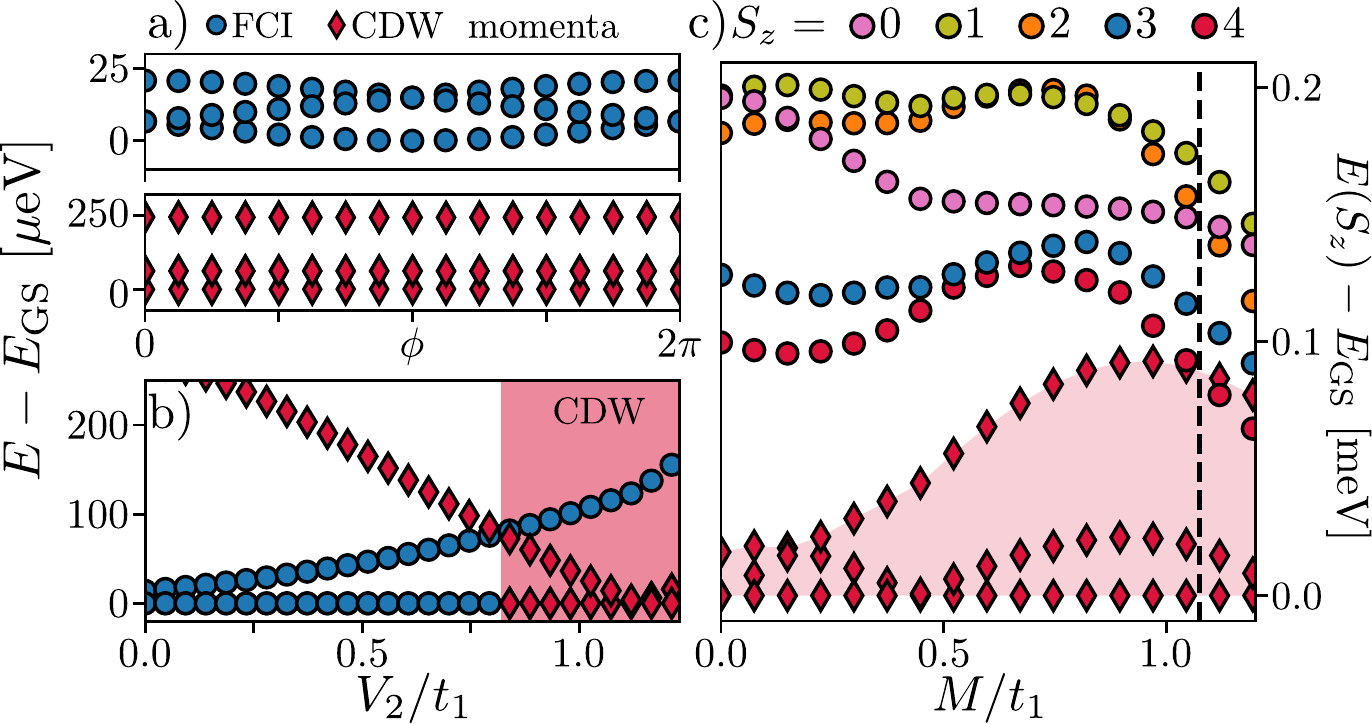}
\caption{a) Spectral flow of the ground state manifold obtained with truncated ED for $V_2=0$ (top) and $V_2 \gg t_1$ (bottom). b) The transition from FQAH to CDW occurs when all CDW ground state are below the highest FCI ground state (all located thanks to their momentum, see text). c) The spread $\delta$ between the three FCI states increases with the applied electric field $M$, until the many-body gap closes and the system forms a ferromagnetic metal. 
}
\label{fig:ElectricFields}
\end{figure}

To study this effect in our model, we include a next-nearest neighbor interaction of strength $V_2$, as discussed in the main text. We provide more details on the characterization of the low-lying states in Fig.~\ref{fig:ElectricFields}a,, where we show their spectral flow. Our results clearly show a transition of the spectral flow from FQAH-like at $V_2=0$ (top) to a insulating one for $V_2 \gg t_1$ (bottom), whose six-fold degeneracy precisely match the one of the $\sqrt{3}\times\sqrt{3}$ CDW.

Let us finally comment on the fate of the FQAH upon application of an electric field. Because the Wannier orbitals at the XM and MX regions are localized on the top and bottom layers, respectively, electric fields materialize as a sublattice potential difference $\mathcal{H}_E = \frac{M}{2} ( \sum_{r\in {\rm XM}} n_r - \sum_{r\in {\rm MX}} n_r )$ in the Hamiltonian. This displacement field appears detrimental to the FQAH as it increases the topmost band's dispersion and tends to localize the Berry curvature non-uniformly near the bottom of the valence band. This intuition is validated by our band-projected ED results, shown in Fig.~\ref{fig:ElectricFields}c, where we observe that the spread between the three spin-polarized FCI states increases with $M$. For the parameters of Fig.~\ref{fig:FirstHintFCI}a, one of the FCI states merges with the many body continuum at $M \simeq 1.04 t_1$, definitely destroying the FQAH phase beyond that point. Note that, as shown App.~\ref{app:continuummodel}, the gap remains at least twice larger than the top valence band's width in all our calculations, justifying our band projected method.

While the FQAH physics disappears upon application of large enough electric fields, we notice that the interaction induced ferromagnetism survives, as it is less sensitive to the changes in the Berry curvature distribution caused by $M$. For instance, the lowest spin-$S_z$ excitation above the fully polarized ground state remains of order $\SI{0.1}{\milli\electronvolt}$ (see Fig.~\ref{fig:ElectricFields}c). This suggests a transition from the FQAH phase to a ferromagnetic metal under increasing electric fields, which could not appear in perfectly flat Landau levels, highlighting the tunability of $t$TMDs. 

\end{document}